\documentclass{SciPost}

\hypersetup{
    colorlinks,
    linkcolor={red!50!black},
    citecolor={blue!50!black},
    urlcolor={blue!80!black}
}

\usepackage[bitstream-charter]{mathdesign}
\DeclareSymbolFont{usualmathcal}{OMS}{cmsy}{m}{n}
\DeclareSymbolFontAlphabet{\mathcal}{usualmathcal}

\fancypagestyle{SPstyle}{
\fancyhf{}
\lhead{\colorbox{scipostblue}{\bf \color{white} ~SciPost Physics Core }}
\rhead{{\bf \color{scipostdeepblue} ~Submission }}

\fancyfoot[C]{\textbf{\thepage}}
}

\usepackage{math_func}
\usepackage{subcaption}
\usepackage{float}

\usepackage[normalem]{ulem}

\newcommand{\placeholderfig}[2][width=\linewidth]{%
  \IfFileExists{#2}{\includegraphics[#1]{#2}}{%
    \fbox{\parbox[c][6cm][c]{0.85\linewidth}{\centering
      \textcolor{red}{\textbf{New figure -- placeholder}\\[1ex]
      upload file: \texttt{\detokenize{#2}}}}}}}

\begin{document}

\pagestyle{SPstyle}

\begin{center}{\Large \textbf{\color{scipostdeepblue}{
Characterizing Mott Insulators\\ in the Interacting One-Body Picture\\
}}}\end{center}

\begin{center}\textbf{
Theo N. Dionne\textsuperscript{1$\star$},
Santiago Villodre\textsuperscript{2,3},
Mikel Iraola\textsuperscript{2,4} and
Maia G. Vergniory\textsuperscript{1,2,5$\dagger$}
}\end{center}

\begin{center}
{\bf 1} Département de Physique et Institut Quantique, Université de Sherbrooke, Sherbrooke, J1K 2R1 Québec, Canada
\\
{\bf 2} Donostia International Physics Center, 20018 Donostia-San Sebastián, Spain
\\
{\bf 3} University of the Basque Country (UPV/EHU), Donostia-San Sebastián, Spain
\\
{\bf 4} Leibniz Institute for Solid State and Materials Research, IFW Dresden, Helmholtzstraße 20, 01069 Dresden, Germany
\\
{\bf 5} Regroupement Québécois sur les Matériaux de Pointe (RQMP), Québec, Canada
\\[\baselineskip]
$\star$ \href{mailto:theo.nathaniel.dionne@usherbrooke.ca}{\small theo.nathaniel.dionne@usherbrooke.ca}\,,\quad
$\dagger$ \href{mailto:maia.vergniory@usherbrooke.ca}{\small maia.vergniory@usherbrooke.ca}
\end{center}

\section*{\color{scipostdeepblue}{Abstract}}
\textbf{\boldmath{%
The one-body picture underlies our understanding of weakly interacting solids but breaks down in strongly correlated systems. We develop a general framework, based on the single-particle Green's function and the one-body reduced density matrix (1RDM), to characterize correlated electronic phases. Applying it to the Hubbard diamond chain, we combine density matrix renormalization group and cellular dynamical mean-field theory to construct symmetry-resolved effective orbitals and track their evolution across its Mott transitions, while the 1RDM purity provides a scalar indicator of the phase boundaries. These tools offer a general route to extend one-body concepts to correlated materials.}}


\vspace{\baselineskip}

\noindent\textcolor{white!90!black}{%
\fbox{\parbox{0.975\linewidth}{%
\textcolor{white!40!black}{\begin{tabular}{lr}%
  \begin{minipage}{0.6\textwidth}%
    {\small Copyright attribution to authors. \newline
    This work is a submission to SciPost Physics Core. \newline
    License information to appear upon publication. \newline
    Publication information to appear upon publication.}
  \end{minipage} & \begin{minipage}{0.4\textwidth}
    {\small Received Date \newline Accepted Date \newline Published Date}%
  \end{minipage}
\end{tabular}}
}}
}


\vspace{10pt}
\noindent\rule{\textwidth}{1pt}
\tableofcontents
\noindent\rule{\textwidth}{1pt}
\vspace{10pt}


\section{Introduction} \label{intro}
The success of band theory stems from the fact that a wide range of crystalline solids can be understood from an effective one-body description. Within this framework, electronic bands, their symmetry representations, and their topology provide a unified language for classifying materials and predicting their physical properties. In recent years, this viewpoint has led to major advances in the understanding and discovery of topological quantum materials and has become an essential component of modern first-principles materials theory.

Strong electronic correlations fundamentally challenge this picture. Electron-electron interactions redistribute spectral weight, generate Hubbard bands, renormalize quasiparticles, and can drive interaction-induced insulating phases that have no counterpart within independent-particle theory. Modern ab initio workflows combining density functional theory and cellular dynamical mean-field theory~\cite{bacq-labreuil_towards_2025} routinely produce fully dressed single-particle Green's function for real materials. The resulting spectral functions provide a direct link to experiment through ARPES. However, a systematic framework for characterizing distinct correlated insulating phases directly out of these single-particle quantities remains elusive. For example, it is known that qualitatively \textit{different} Mott insulators can be induced on the square lattice~\cite{yao_fragile_2010}, yet no model-agnostic diagnostic based purely on the Green's function exists to distinguish them. 

Two complementary ideas naturally suggest such a framework. Crystal symmetry provides a powerful organizing principle throughout condensed matter \cite{bradlyn_topological_2017, Altland_Simons_2010}. In particular, the irreducible representations of the little group at high-symmetry points label the eigenstates of the non-interacting problem. Moreover, as we make precise in section~\ref{sec:theory}, these labels extend to the interacting Green's function as well. In the orbital basis, constructing effective one-body orbitals from the one-body reduced density matrix (1RDM) has been shown to expose the dominant microscopic degrees of freedom at low energy~\cite{verma_local_2025}. 

These complementary quantities define what we call the interacting one-body framework, a systematic approach for characterizing correlated insulators using experimentally and computationally accessible single-particle quantities. Starting from the interacting single-particle Green's function, the framework combines three complementary levels of information. First, the spectral function is decomposed into the irreducible representations of the little group at high-symmetry points, providing symmetry-resolved information on the single-particle excitations. Second, the one-body reduced density matrix (1RDM), obtained directly from the Green's function, captures the average charge distribution of the many-body ground state through its effective one-body orbitals. Finally, the purity of the 1RDM provides a simple scalar measure of the deviation from a Slater determinant, offering a compact diagnostic of correlation-driven phase transitions.

As a proof of principle, we apply this method to the Hubbard diamond chain~\cite{iraola_towards_2021, soldini_interacting_2023}. This one-dimensional model hosts three distinct correlated phases driven by the interplay of Hubbard interactions and spin-orbit coupling. We first confirm the existence of these phases using tensor network (DMRG) calculations. We then fix the interaction strength and sweep across the phases, characterizing each one through symmetry-resolved spectral functions, effective one-body orbitals, and the purity of the one-body reduced density matrix. In particular, the purity of the 1RDM proves to be discontinuous at the transition between the Mott insulator and the spin-orbit-induced atomic insulator, providing a sharp single-particle diagnostic of a transition that is otherwise difficult to resolve from the spectral function alone. More generally, our results establish a transferable framework for characterizing correlated electronic phases directly from interacting single-particle quantities.

The remainder of the paper is organized as follows. In Section~\ref{sec:theory}, we introduce the theoretical ingredients of the interacting one-body framework: the symmetry constraints on the single-particle Green's function and the one-body reduced density matrix. In Section~\ref{subsec:theory_model}, we introduce the Hubbard Diamond Chain model, discuss its correlated phases, and derive the microscopic origin of the symmetry-allowed spin--orbit coupling term. Section~\ref{methods} presents the numerical methods, with DMRG used to determine the many-body phase diagram and CDMFT used to compute the interacting Green's function. In Section~\ref{results_discussion}, we apply the interacting one-body framework to the Hubbard Diamond Chain by analyzing symmetry-resolved spectral functions, effective one-body orbitals, and the purity of the one-body reduced density matrix. Finally, Section~\ref{conclusions} summarizes our results and discusses the broader applicability of the framework to correlated quantum materials.



\section{Theory}\label{sec:theory}

We now present the theoretical foundations of the interacting one-body framework. Subsection~\ref{subsec:theory_SPGF_symmetry} establishes the symmetry constraints on the single-particle Green's function (SPGF), while subsection~\ref{subsec:onerdm} introduces the one-body reduced density matrix (1RDM) and the effective one-body orbitals and purity derived from it.

\subsection{Symmetry of the SPGF}\label{subsec:theory_SPGF_symmetry}

In this section, we demonstrate that the symmetries of the system generate a unitary representation of the corresponding group, which in turn imposes rigorous constraints on the structure of the Green’s function. Upon incorporating crystalline symmetries, we explicit how the spectral function can be decomposed into symmetry sectors corresponding to the irreps of the little group.

\subsubsection{Wigner's theorem}\label{subsubsec:theory_Wigner}

Consider a unitary transformation $\hat{\mathrm{U}}$ in the $N$-particle Hilbert space $\mathcal{H}^{(N)}$ which has a matrix representation $U_{\mu\nu}$ on the set of fermionic creation and annihilation operators as \cite{gurarie_single-particle_2011, lessnich_elementary_2021}:
\begin{align}
    \hat{\mathrm{U}}\hat{\mathrm{c}}_\mu \hat{\mathrm{U}}^\dagger &= \sum_{\nu}U_{\mu\nu}\hat{\mathrm{c}}_\nu &
    \hat{\mathrm{U}}\hat{\mathrm{c}}_\mu^\dagger \hat{\mathrm{U}}^\dagger &= \sum_{\nu}U_{\nu\mu}^*\hat{\mathrm{c}}_\nu^\dagger \label{eq:ladders_under_unitaries}
\end{align}
One can see that the complex frequency Green's function is invariant under the matrix transformation defined above as long as it is a symmetry of the many-body system (see Appendix \ref{annex:symmetry_SPGF}). Explicitly,
\begin{align}
    \sum_{\alpha\beta}U_{\mu\alpha}G_{\alpha\beta}(z)U_{\beta\nu}^* = G_{\mu\nu}(z)\label{eq:sym_garochée}
\end{align}
It is known \cite{liubarskii_application_1960} that if $\mathrm{U}$ is a representation of a finite group, then one can find a basis such that:
\begin{align}
    \mathrm{U}(g) = \bigoplus_i \mathbbm{1}_{m^i}\otimes\mathrm{U}^i(g)
\end{align}
with $m^{i}$, the multiplicity of irrep $i$. Since the Green's function in matrix form commutes with the full set of representation matrices \eqref{eq:sym_garochée}, it can be block diagonalized into symmetry sectors upon using Schur's lemma and changing basis \cite{MikelPhD, liubarskii_application_1960}:
\begin{align}
    \mathrm{G}(z) = \bigoplus_i\mathbbm{1}_{d^i}\otimes\mathrm{G}^{i}_{m^i\times m^i}(z)
\end{align}
where $d^i$ is the complex dimension of the associated irrep. This form indicates that the Green's function possesses $m^i$ generally distinct $d^i$-degenerate eigenvalues identified by the irrep $(i)$. This decomposition is valid for any complex frequency $z\in\C$.

\subsubsection{Application to crystalline space groups} \label{subsubsec:theory_crys_sym}

In the case of a lattice, its symmetry group $\mathscr{G}$ is called the space group \cite{MikelPhD}. A notable subgroup of $\mathscr{G}$ is $T\subset \mathscr{G}$, the crystalline translation subgroup. Applying the results of the previous section (\ref{subsubsec:theory_Wigner}) yields:
\al{
    G(z) \sim \bigoplus_{\mathbf{k}\in\text{1BZ}}G(\mathbf{k}, z)
}
which is how one denotes the common notion that crystalline translation invariance yields a SPGF as a function of crystal momentum in group theoretic terms. Within the subspace belonging to a given wavevector $\mathbf{k}$, the remaining symmetry group is called the little group $\mathscr{G}_\mathbf{k}$ and is defined as \cite{MikelPhD}:
\al{
    \mathscr{G}_\mathbf{k} = \{g\in \mathscr{G} ~|~ g\mathbf{k} = \mathbf{k}~(\text{mod}~\mathbf{G}\in\text{recip. latt.})\}
}
The irreps of these groups offer strong symmetry-based labels for the eigenvectors and eigenvalues of the Green's function off the basis of crystal symmetry. In practice, all irreps of the little groups for every space group are available on the Bilbao Crystallographic Server \cite{BSC}.


\subsection{One-body reduced density matrix}\label{subsec:onerdm}

Here, we show how the one-body reduced density matrix (1RDM) can be used as a tool for analyzing average behaviour in a many-body system at the one-particle level. In particular, a scheme for analyzing average orbital charge distribution is laid out with a short introduction to the use of purity in the context of 1RDMs.

\subsubsection{Generalities}\label{subsubsec:theory_onerdm_generalities}

The exact N-body density matrix for a system of interacting particles is generally impossible to calculate. However, some information about the system can still be obtained from the 1RDM \cite{solovej2014manybody}. Following \cite{gross_many-particle_1991}, the 1RDM is defined as the partial trace of the full density matrix over the degrees of freedom relative to $N-1$ particles:
\begin{align}
    \gamma_{\mu\nu} = \bra{\mu}\operatorname{Tr}_{N-1}\left\{\hat{\rho}^{(N)}\right\}\ket{\nu}
\end{align}
where the Greek indices label the remaining degrees of freedom. It can be shown (as done in Appendix \ref{annex:1RDM}) that this simply reduces to the expectation value of a pair of creation and annihilation operators. In turn, the expectation value of a one-body term can be computed directly from the complex frequency Green's function \cite{rickayzen_greens_2013, dionne_pyqcm_2023}:
\begin{equation}
    \gamma_{\mu\nu} = \left\langle\hat{\mathrm{c}}_\nu^\dagger\hat{\mathrm{c}}_\mu\right\rangle \overset{T=0}{=} \oint_{\mathcal{C}_<}\frac{\text{d}z}{2\pi i}G_{\mu\nu}(z)\label{eq:gamma_from_G}
\end{equation}
The above relation is valid at zero temperature. In the case of this work, we readily have access to $\gamma_{a\sigma,b\sigma'}(\mathbf{k})$ where latin indices indicate orbitals and $\sigma(\sigma')$ denote spin. It can be checked that the filling of the model is related to the trace of the 1RDM
\begin{align}
    n = \frac{1}{N_{\text{DoF}}}\sum_{\mathbf{k}\in\text{B.Z.}}\text{Tr}\{\boldsymbol{\gamma}(\mathbf{k})\} = \frac{1}{N_{\text{DoF}}}\sum_{\mathbf{k}\in\text{B.Z.}}\sum_{a\sigma}\left\langle\mathrm{n}_{a\sigma}(\mathbf{k})\right\rangle
\end{align}
where $N_{\text{DoF}}$ is the number of lattice degrees of freedom. In general, all one-body operator expectation values can be obtained from the 1RDM \cite{dionne_pyqcm_2023}
\begin{align}
    \left\langle\mathrm{t}\right\rangle = \left\langle\mathrm{c}_{\mu}^\dagger t_{\mu\nu}\mathrm{c}_{\nu}\right\rangle = \text{Tr}\{\boldsymbol{t}\boldsymbol{\gamma}\}
\end{align}

\subsubsection{Effective one-body orbitals}\label{subsubsec:theory_onerdm_orbitals}

In order to study the orbital distribution of charge in an N-body system, it is possible to trace out all degrees of freedom except for orbital and spin indices. In this case, $\gamma$ can be expressed in terms of its eigenvectors and eigenvalues:
\begin{align}
    \gamma = \sum_{j}p_j\ket{\phi_j}\!\!\bra{\phi_j}
\end{align}
where $p_j\in[0,1]$ are classical occupations and $\ket{\phi_n}$ are effective one-particle orbitals.

\subsubsection{Density matrix purity}\label{subsubsec:purity}

Finally, it is intuitively clear that in the case of a \textit{Slater determinant} type ground state
\begin{align}
    \ket{\Psi^{(N)}_{0}} = \bigotimes_{\mathbf{k}\in \text{B.Z.}}\ket{\psi_{0}(\mathbf{k})}
\end{align}
tracing out all wave-vectors except $\mathbf{k}$ will yield a pure state for the resulting 1RDM. In fact, it is known that \cite{gross_many-particle_1991}:
\begin{align}
    \gamma^2 = \gamma ~\Leftrightarrow~ \ket{\Psi^{(N)}_0}\text{ is a Slater determinant}
\end{align}
Therefore, the purity $\text{Tr}\{\gamma^2\}$ acts as a measure of electronic interactions.

The 1RDM and the Green's function are complementary rather than redundant. The Green's function offers frequency-resolved information about the symmetry of the problem, acting mainly as a proxy for irrep-labelled bands in the presence of interactions. Conversely, the 1RDM is derived directly from the Green's function through equation~\eqref{eq:gamma_from_G}, integrating out the frequency information in the process. Hence, the 1RDM represents the occupied spectral weight, thus representing a static yet orbital-resolved single-particle reduction of the full many-body ground state.


\section{Model}\label{subsec:theory_model}

In this section, we introduce the Hubbard Diamond Chain (HDC) model and derive the most general symmetry-allowed spin-orbit coupling (SOC) term compatible with its crystal symmetries.

\subsection{Hubbard Diamond Chain}\label{subsubsec:theory_HDC_model}

\begin{figure}[h]
    \centering
    \includegraphics[width=0.7\linewidth]{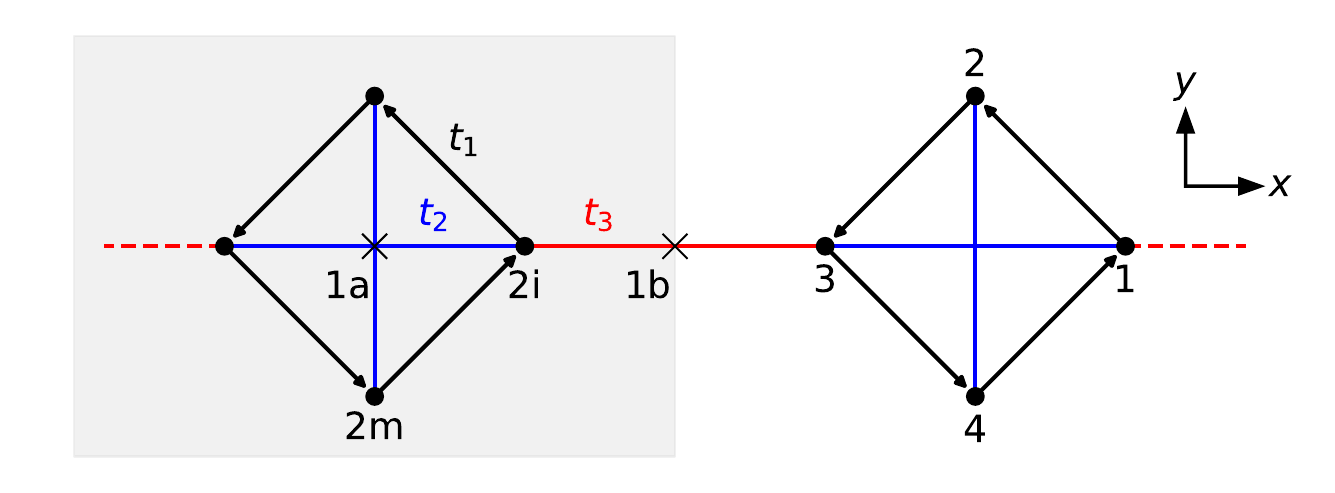}
    \caption{The Hubbard diamond chain. Each site contains one spinful s-like orbital.}
    \label{fig:HDC_chain}
\end{figure}

The Hubbard Diamond Chain (HDC) considered in this work consists of a one-dimensional array of diamond-shaped clusters \cite{iraola_towards_2021, soldini_interacting_2023} (Fig.~\ref{fig:HDC_chain}). It can be regarded as the one-dimensional analogue of the lattice introduced in Ref.~\cite{yao_fragile_2010}.

Through the use of symmetry arguments (Appendix \ref{annex:sym_analysis}), the tight binding hamiltonian for this lattice is taken to be:
\begin{align}
    \hat{\mathrm{H}} = - \sum_{i,ab,\sigma}\hat{\mathrm{c}}_{i,a,\sigma}^\dagger\mathbb{T}_{ab}^{\sigma}\hat{\mathrm{c}}_{i,b,\sigma} -t_3\sum_{i,\sigma}\left(\hat{\mathrm{c}}_{i,1,\sigma}^\dagger\hat{\mathrm{c}}_{i\!+\!1,3,\sigma}+\text{h.c.}\right)+U\sum_{i,a}\hat{\mathrm{n}}_{i,a,\uparrow}\hat{\mathrm{n}}_{i,a,\downarrow} \label{eq:HDC_Hamiltonian}
\end{align}
with $i$ indicating the unit cell, $a(b)$ the orbital in the unit cell and $\sigma$ the electron's spin. The intra-diamond coupling matrix is explicitly given by:
\begin{align}
    \mathbb{T}_{ab}^{\sigma} = \begin{pmatrix}
        0 & t_1\Exp{i\sigma\phi} & t_2 & t_1\Exp{-i\sigma\phi}\\
        t_1\Exp{-i\sigma\phi} & 0 & t_1\Exp{i\sigma\phi} & t_2\\
        t_2 & t_1\Exp{-i\sigma\phi} & 0 & t_1\Exp{i\sigma\phi}\\
        t_1\Exp{i\sigma\phi} & t_2 & t_1\Exp{-i\sigma\phi} & 0
    \end{pmatrix}\label{eq:hopping_matrix}
\end{align}
where $\sigma = \{+1,-1\}$ for $\{\uparrow,\downarrow\}$.

Having established the microscopic Hamiltonian, we now discuss the symmetry properties and correlated phases that will serve as the benchmark for the interacting one-body framework. Although the HDC is a one-dimensional model, it can be embedded in a three-dimensional crystal by considering one extended lattice direction and two transverse directions with a single-unit-cell periodicity. This allows the symmetry of the model to be described by the three-dimensional space group $Pmmm$ (No.~47).

We place one spinful spatially isotropic orbital per site in the lattice. Given their positions within the unit cell (Wyckoff positions 2i and 2m), we can use the Bilbao Crystallographic Server \cite{bradlyn_topological_2017, vergniory_graph_2017, elcoro_double_2017} to obtain the following double little group irreps at high symmetry points:
\begin{align}
    \left(\bar{E}_g\uparrow G\right)_{2i} \oplus \left(\bar{E}_g\uparrow G\right)_{2m} : \left\{2\bar{\Gamma}_5\oplus2\bar{\Gamma}_6, 2\bar{X}_5\oplus2\bar{X}_6\right\}
\end{align}

As seen in section \ref{subsec:theory_SPGF_symmetry}, the irreps obtained are also valid for the complex Green's function. Previous work on the model \cite{MikelPhD} showed that for a single diamond with $t_2/|t_1|= 0.5$, three distinct phases emerge as a function of $\phi$ and $U$, as determined through exact diagonalization (ED). At values around $\phi=0$ and $\phi=\pi/2$, the phases are Mott insulators and are named Mott-I and Mott-II, respectively. These phases arise from two distinct metallic phases in the non-interacting model as presented in Appendix~\ref{annex:non_interacting_spectral}. At intermediary values of SOC, the insulating phase of the model is not a Mott insulator, but rather a phase adiabatically connected to a band insulator whose gap is opened by SOC rather than by Coulomb repulsion. Since the non-interacting phase is named the Spin-orbit induced Atomic Insulator (SAI), the related interacting phase is named the SAI+U.

As done in \cite{MikelPhD}, the phase boundary can be determined through two avenues. On one hand, the many-body gap closes at the phase transition as the lowest energy eigenvalue exchanges with the next lowest. On the other hand, the expectation value of the mirror operator along the $x$-direction remains $\langle M_x\rangle = 1$ in the case of the SAI+U, since the phase is adiabatically connected to a Slater determinant state in which $\langle M_x\rangle = 1$ always holds. Conversely, any phase with $\langle M_x\rangle = -1$ cannot be adiabatically connected to a Slater determinant. This is the case for Mott-I and Mott-II.

In the present work, we extend this analysis using Tensor Network methods to verify the previous findings and to explore whether the same phase structure persists for $t_2/|t_1|= 0.8$.

Although the form of the SOC term is fully determined by symmetry (Appendix~\ref{annex:sym_analysis}), we derive it microscopically in the following section to give physical interpretation to the parameter $\phi$ and to establish that its non-redundant range is $\phi\in[0,\pi/2]$.




\subsection{Microscopic origin of the SOC term}\label{subsec:theory_microscopic_SOC}

Although the SOC $\phi$ parameter is \textit{allowed} by symmetry (cf. Appendix \ref{annex:sym_analysis}), it can be physically motivated by constructing the tight-binding model in a bottom-up approach. The physically relevant microscopic elements for deriving the form of the hopping terms are the kinetic and the SOC contributions to the Hamiltonian of a free electron \cite{vanderbilt2018berry}:
\begin{align}
    \mathrm{H} = \mathrm{H}_\text{kin} + \mathrm{H}_{\text{SOC}} = \frac{\mathbf{p}^2}{2m} + \kappa\boldsymbol\sigma\cdot\boldsymbol{\nabla}U(\mathbf{r})\times\mathbf{p}\label{eq:free_terms}
\end{align}
In order to construct the tight-binding basis, we choose a unit cell index $i$, an orbital index $a(b)$ and a spin index $\sigma$. Then, the matrix elements of these terms \eqref{eq:free_terms} are evaluated in the basis of the physical orbitals of the model. Here, we assume that every lattice site represents a positively charged ion with a perfectly radial charge potential $U(r)$. Furthermore, we only consider real s-like electronic orbitals based around every ion.

Starting with the kinetic term, the matrix elements can be written and evaluated as:
\begin{align}
    t^{\text{kin}}_{ia\sigma;jb\sigma'} &= \bra{ia\sigma}\frac{\mathbf{p}^2}{2m}\ket{jb\sigma'} = \delta_{\sigma\sigma'}\frac{-\hbar^2}{2m}\int\text{d}^3r\, w_{ia}(\mathbf{r})\nabla^2w_{jb}(\mathbf{r})
\end{align}
As expected, $\bra{ia\sigma}\frac{\mathbf{p}^2}{2m}\ket{jb\sigma'} = \bra{jb\sigma'}\frac{\mathbf{p}^2}{2m}\ket{ia\sigma}$ in the case of real orbitals and thus $t^{\text{kin}}_{ia\sigma;jb\sigma'}\in\mathbb{R}$.

Now, for the SOC term, the matrix elements are slightly more constrained by the geometry of the system. As a matter of fact, since both the gradient of the electrostatic potential and the average electron momentum is odd out-of-plane, only the $\sigma_z$ term has any chance of surviving. Therefore,
\begin{align}
    t^{\text{SOC}}_{ia\sigma;jb\sigma'} &= \bra{ia\sigma}\kappa\sigma_z\hat{\mathbf{z}}\cdot\boldsymbol{\nabla}U(\mathbf{r})\times\mathbf{p}\ket{jb\sigma'}\notag\\ &= -i\hbar\kappa\bra{\sigma}\sigma_z\ket{\sigma'}\hat{\mathbf{z}}\cdot\int\text{d}^3r\, w_{ia}(\mathbf{r})\boldsymbol{\nabla}U(\mathbf{r})\times\boldsymbol{\nabla}w_{jb}(\mathbf{r})
\end{align}
In a similar fashion than for the kinetic term, it can be observed that the SOC term is purely imaginary. However, the geometry of the HDC renders the SOC contributions to $t_2$ and $t_3$ trivial since the gradient of the electrostatic potential and the gradient of the s-like orbitals are both spatially odd with respect to Wyckoff positions 1a and 1b. So, in the spin basis
\begin{align}
    \tilde{t}_2 &= t_2\mathbbm{1} & \tilde{t}_3 &= t_3\mathbbm{1}
\end{align}

The relevant terms which do not have a geometric cancellation are the diagonal links pertaining to $t_1$. Since the kinetic term is purely real and the SOC term is purely imaginary, we can write:
\begin{align}
    \tilde{t}_1 &= t_1^{\text{kin}}\mathbbm{1} + t_1^{\text{SOC}}\sigma_z
\end{align}
Focusing on the spin up sector, the contribution can be parametrized by
\begin{align}
    t_1^{\text{kin}} + t_1^{\text{SOC}} = t_1\text{e}^{i\phi}
\end{align}
where
\begin{align}
    t_1 &= \sqrt{(t_1^{\text{kin}})^2 + |t_1^{\text{SOC}}|^2} & \phi &= \text{arctan}\left(\frac{|t_1^{\text{SOC}}|}{t_1^{\text{kin}}}\right)\label{eq:diagonal_hopping_term_definition}
\end{align}
which allows us to make contact with the form derived via symmetry:
\begin{align}
    \tilde{t}_1=\begin{pmatrix}
        t_1\text{e}^{i\phi} & 0\\0 & t_1\text{e}^{-i\phi}
    \end{pmatrix}
\end{align}
Furthermore, the SOC phase parameter $\phi$ can be further interpreted by considering an electron circulating along a closed loop around a diamond in the chain (cf. fig. \ref{fig:HDC_chain}).

On one hand, hopping counterclockwise on the diamond acts as $\Exp{i\phi}\sigma_z$ in spin space \eqref{eq:HDC_Hamiltonian}, showing already that the phase winds in opposite directions for each spin projection as imposed by the SOC term containing $\mathrm{L}\cdot\mathrm{S}$. Moreover, since the chain is confined to the $xy$ plane, it only has orbital angular momentum along $\hat{\mathrm{z}}$, thus justifying \textit{why} there is no spin mixing in this chain.

On the other hand, the total phase accumulated from a full rotation will be $\Exp{\pm 4i\phi}$, hence, the non-redundant phase angles are contained within $\phi\in[0,\pi/2]$ due to angles being equivalent modulo $2\pi$. It follows that the effects of spin orbit coupling are maximal when $\phi = \pi/2$ which corresponds naturally with the definition of the angle \eqref{eq:diagonal_hopping_term_definition}.

\section{Methods} \label{methods}

The interacting one-body framework introduced in Section~\ref{sec:theory} relies on two complementary many-body approaches. We will employ DMRG to establish the many-body phase diagram, providing an independent benchmark against which the framework can be validated. CDMFT, in turn, gives direct access to the interacting single-particle Green's function, which constitutes the starting point of the interacting one-body framework. Together, these methods allow us to assess whether the correlated phases identified from many-body observables can be faithfully characterized using only single-particle quantities.

\subsection{Tensor Networks} \label{subsec:methods_TN_tenpy}

As a benchmark for the interacting one-body framework, we first determine the many-body phase diagram using Tensor Network (TN) methods. TN methods provide a powerful and efficient framework for simulating quantum many-body systems \cite{Or_s_2014, Ba_uls_2023}. Among them, the Density Matrix Renormalization Group (DMRG) is the most widely employed variational algorithm for studying low-dimensional strongly correlated systems \cite{Schollw_ck_2011, Verstraete_2008}.

In this work, we use DMRG to identify the three distinct phases reported in \cite{MikelPhD} for the model introduced in section~\ref{subsec:theory_model}. The simulations were performed using the TenPy Tensor Network library for Python~\cite{tenpy2018, Hauschild_2018}, a dedicated platform for TN simulations in condensed matter physics. Within this framework, the lattice and Hamiltonian were explicitly constructed as a Matrix Product Operator (MPO), while the many-body wavefunction was represented as a Matrix Product State (MPS) optimized through the two-site DMRG algorithm.

To determine the energy gap, we first compute a well-converged approximation to the ground-state wavefunction and subsequently perform a second DMRG calculation in which the new target state is explicitly constrained to be orthogonal to the ground state. The energy difference between these two states then provides the excitation gap of the system.

The expectation value $\langle M_x\rangle$ of the mirror operator is calculated by applying a site permutation to the ground-state MPS, followed by a spin-flip transformation using the $S_x$ operator. The overlap between this transformed MPS and the original wavefunction yields the desired expectation value.

This procedure is systematically repeated across a grid of $U$ and $\phi$ values to generate the data shown in figure~\ref{fig:TN_unified}. Further technical details on the Tensor Network simulations are provided in Appendix~\ref{annex:TN_details}.

While the single-particle Green's function is in principle also accessible within DMRG~\cite{Schollw_ck_2011, hallberg_density_1999}, its computation is considerably more demanding than that of ground state observables such as the energy gap and $\langle M_x\rangle$. The Green's function is instead the primary output of Cluster Dynamical Mean Field Theory, which naturally motivates the complementary use of both methods in this work.

\subsection[Cluster dynamical mean-field theory]{\texorpdfstring{ Cluster dynamical mean-field theory}{Cluster dynamical mean-field theory}}\label{subsec:methods_CDMFT}

Cluster dynamical mean-field theory (CDMFT) is the cluster extension of the well known dynamical mean-field theory \cite{GeorgesDMFT, KotliarCDMFT}. This method is typically used when local physics dominates the overall behaviour of the system. Although typically associated with higher-dimensional systems, CDMFT has been successfully applied to one-dimensional models, including the Hubbard chain~\cite{bolech_cellular_2003}, where comparisons with DMRG confirm that the method captures local single-particle quantities accurately. It can also be viewed as a self-consistent extension to cluster perturbation theory (CPT) \cite{dionne_pyqcm_2023} which improves the Green's function of the cluster by coupling it to a non-interacting bath configured to mimic the presence of the remaining lattice beyond the cluster on a mean-field level. A pedagogical review of the theory underpinning both methods is found in \cite{pavarini_many-body_2015}.

In this work, CDMFT is used to obtain the Green's function of the HDC model presented in section~\ref{subsec:theory_model}. All calculations are performed using the pyqcm library~\cite{dionne_pyqcm_2023}. Within pyqcm, the tight-binding model is constructed using the utilities provided and an exact diagonalization (ED) solver is used to solve the CDMFT impurity problem.

The general workflow of the CDMFT algorithm proceeds as follows. First, the lattice is re-tiled with clusters and a superlattice such that every site in the original lattice remains. This step can be seen as a generalization of the choice of correlated orbital in DMFT. Then, the cluster is used to construct an effective impurity model: the cluster is coupled to a finite and discrete non-interacting bath by hybridization terms. Next, the entire impurity model is solved using ED to obtain the Green's function on the cluster. The lattice Green's function is obtained using Dyson's equation:
\begin{align}
    G^{-1}(\omega, \mathbf{k}) = G_0^{-1}(\omega, \mathbf{k}) - \Sigma_{\text{clus}}(\omega),
\end{align}
with $\Sigma_{\text{latt}}(\omega, \mathbf{k})\approx\Sigma_{\text{clus}}(\omega)$. At this step, the difference between the cluster's Green's function and the projected lattice Green's function is quantified on the imaginary axis using the distance function:
\begin{align}
    d = \sum_{i\omega_n,\mu,\nu}W_n\left|\bar{G}_{\mu\nu}(i\omega_n) - G_{\text{clus},\mu\nu}(i\omega_n)\right|,
\end{align}
with $\bar{G}(\omega) = \sum_{\mathbf{k}}G_{\text{latt}}(\omega, \mathbf{k})$ and $W_n$ being weights assigned to each $i\omega_n$. Comparison is drawn along the imaginary axis as the poles of the Green's function make numerical differences difficult to minimize. The distance function is minimized by varying only the hybridization strengths and bath energies. The optimal parameters are used to construct the next impurity model. The above is repeated until the distance function converges relative to the successive iteration. Details regarding the CDMFT calculations are provided in Appendix~\ref{annex:pyqcm_details}.


\section{Results and Discussion}\label{results_discussion}

Having established the interacting one-body framework and the computational methodology, we now investigate whether the correlated phases can be identified using only single-particle quantities. We first use the essentially exact DMRG method to determine the phase transitions from the many-body excitation gap and the mirror eigenvalue of a single diamond, thereby establishing the reference phase diagram. We then use CDMFT to examine whether the same phase structure can be recovered from the interacting Green's function and its derived one-body quantities in the full Hubbard Diamond Chain.

The results in Section~\ref{subsec:results_TN} are obtained using DMRG, while those in Sections~\ref{subsec:results_spectral}, \ref{subsec:results_one_body_orbitals}, and \ref{subsec:results_purity} are obtained using the CDMFT scheme described in Section~\ref{subsec:methods_CDMFT}.

\subsection{Tensor Network phase calculations}\label{subsec:results_TN}
\begin{figure}[h]
    \centering
    \includegraphics[height=0.29\linewidth]{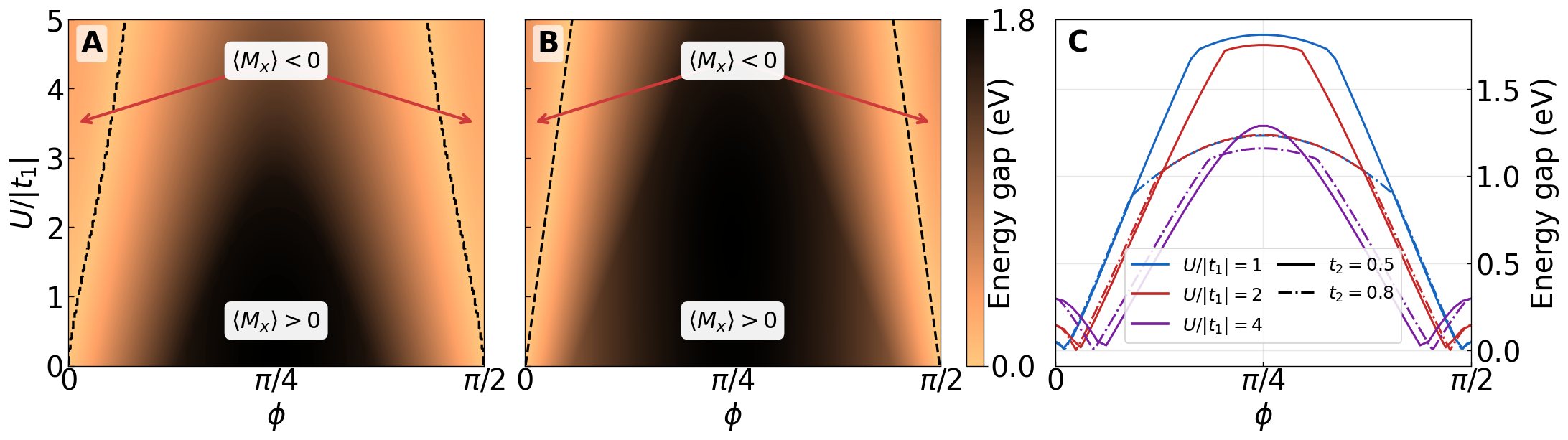}
    \caption[Energy gap for a single diamond (3 panels).]{Energy gap between ground state and the first excited state for a single diamond ($N=1$ unit cell, $t_3=0$, 4 electrons at half-filling). \textbf{(A)--(B)} Energy gap as a function of $U/|t_1|$ and $\phi$ for $t_2/|t_1| = 0.5$ and $t_2/|t_1| = 0.8$, respectively. The dotted lines mark the sign change of $\langle M_x \rangle$, which coincides with the closing of the gap: the central region exhibits $\langle M_x \rangle > 0$, while the lateral regions exhibit $\langle M_x \rangle < 0$. \textbf{(C)} Energy gap as a function of $\phi$ for selected values of $U/|t_1|$ and $t_2/|t_1|$.}
    \label{fig:TN_unified}
\end{figure}

In this section, we first investigate the phase transitions of the single diamond through the expectation value of the mirror operator and the many-body excitation gap. This simple setting provides a clear physical picture of the different phases before turning to their characterization using single-particle quantities. To this end, using DMRG, we compute the many-body excitation gap together with the expectation value of the mirror operator for a single diamond ($N=1$, $t_3=0$) at $t_2/|t_1|=0.5$ and $t_2/|t_1|=0.8$, as shown in figure~\ref{fig:TN_unified}.

Figs.~\ref{fig:TN_unified}(A) and (B) show the excitation gap as a function of $U$ and $\phi$; the dotted lines mark where the mirror expectation value $\langle M_x\rangle$ changes sign, with the sign of $\langle M_x\rangle$ indicated directly on each region of the panels. Three distinct regions are clearly identified. The central trapezoidal region is characterized by a finite excitation gap and a positive mirror expectation value, $\langle M_x\rangle > 0$, corresponding to the SAI+U phase. In contrast, the regions near $\phi = 0$ and $\phi = \pi/2$ remain gapped but exhibit $\langle M_x\rangle < 0$, identifying the Mott-I and Mott-II phases, respectively. The dotted lines therefore trace the phase boundaries: there, the excitation gap closes and $\langle M_x\rangle$ changes sign simultaneously, demonstrating that the change in the mirror eigenvalue is accompanied by a genuine quantum phase transition, as predicted in~\cite{MikelPhD}. Furthermore, as the interaction strength $U$ increases, the transition region broadens, indicating that stronger correlations progressively dominate over the spin--orbit coupling in determining the phase diagram. These results reproduce the phase structure reported in Ref.~\cite{MikelPhD} for $t_2/|t_1| = 0.5$ and demonstrate that it remains robust at $t_2/|t_1| = 0.8$.

The primary effect of increasing $t_2/|t_1|$ is to shift the phase boundaries within the parameter space. To quantify this effect, we fix $U = 4$, the interaction strength used throughout the following sections. At this interaction strength, the phase transitions occur at $\phi = 0.057\pi$ and $\phi = 0.443\pi$ for $t_2/|t_1| = 0.5$, while for $t_2/|t_1| = 0.8$ they shift slightly to $\phi = 0.046\pi$ and $\phi = 0.453\pi$. In contrast, the inter-diamond hopping $t_3$ leaves these critical values of $\phi$ unchanged. As shown in Appendix~\ref{annex:t3_effects}, activating $t_3$ uniformly reduces the excitation gap throughout the phase diagram without modifying the phase boundaries, demonstrating that the location of the phase transitions is governed solely by the intra-diamond physics.

The dependence of the critical values of $\phi$ on $t_2/|t_1|$ reflects the competition between the Hubbard interaction and the SOC-induced insulating state. As $t_2/|t_1|$ increases from $0.5$ to $0.8$, the Mott–SAI+U transition at $U = 4$ shifts to lower values of $\phi$, indicating that a weaker SOC is sufficient to stabilize the SAI+U phase. This behavior can be understood by considering the underlying insulating mechanisms. A Mott insulator originates from the suppression of charge fluctuations by the Hubbard interaction, whereas the SAI+U phase is adiabatically connected to a band insulator, whose gap is generated by the electronic band structure and further stabilized by SOC~\cite{NFMott}. Increasing $t_2$ enhances the non-interacting gap, thereby reducing the additional SOC required to drive the transition into the SAI+U phase.

These DMRG results provide a reference phase diagram against which the CDMFT-based single-particle analysis can be compared. Since CDMFT is not employed here to locate phase boundaries but rather to characterize the single-particle properties deep within each phase, the CDMFT calculations are performed at representative points well within each phase, deliberately avoiding the transition regions.

\subsection{Symmetry labelled spectral function}\label{subsec:results_spectral}

\begin{figure}[htb!]
    \centering
    \begin{subfigure}[t]{0.45\linewidth}
        \centering
        \includegraphics[width=\linewidth]{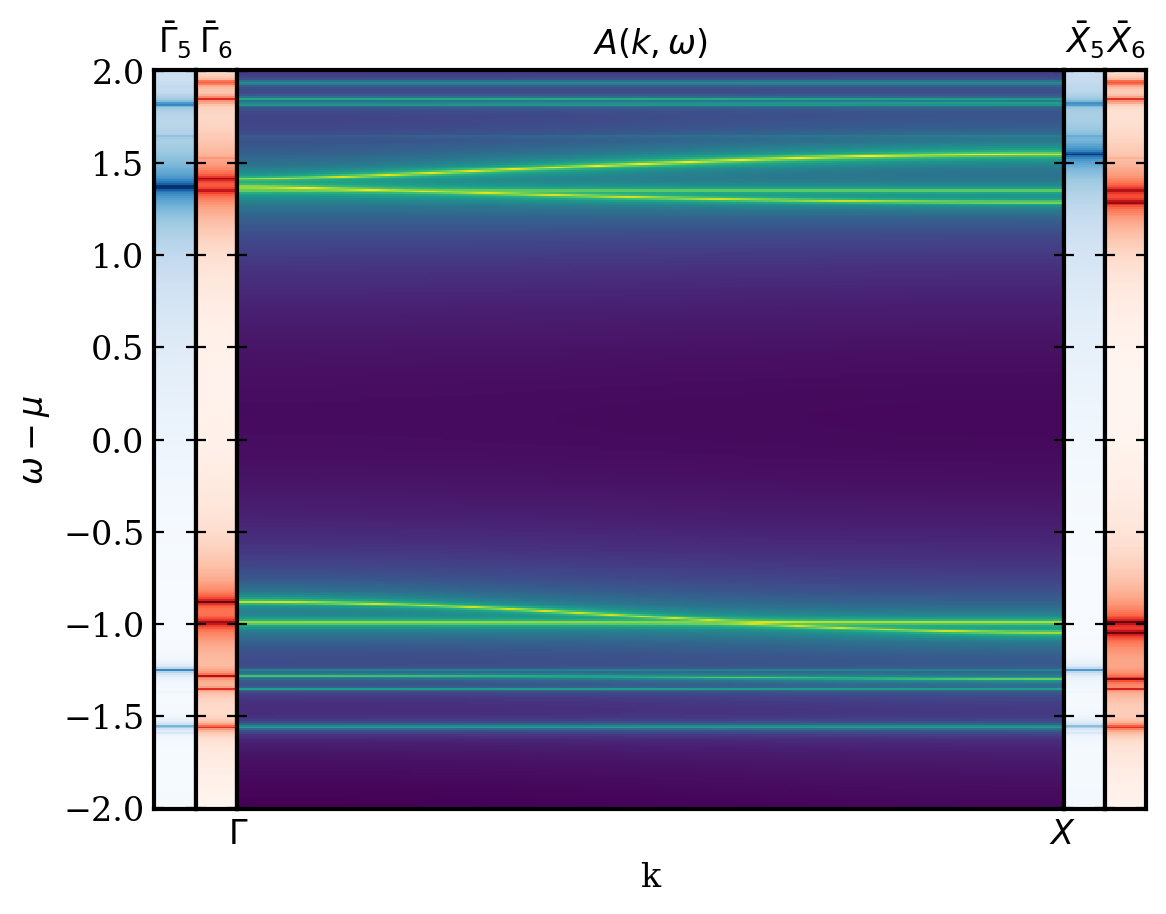}
        \caption{Mott-I at $\phi=10^{-4}\pi$.}
        \label{fig:MottI_spectral}
    \end{subfigure}
    \begin{subfigure}[t]{0.45\linewidth}
        \centering
        \includegraphics[width=\linewidth]{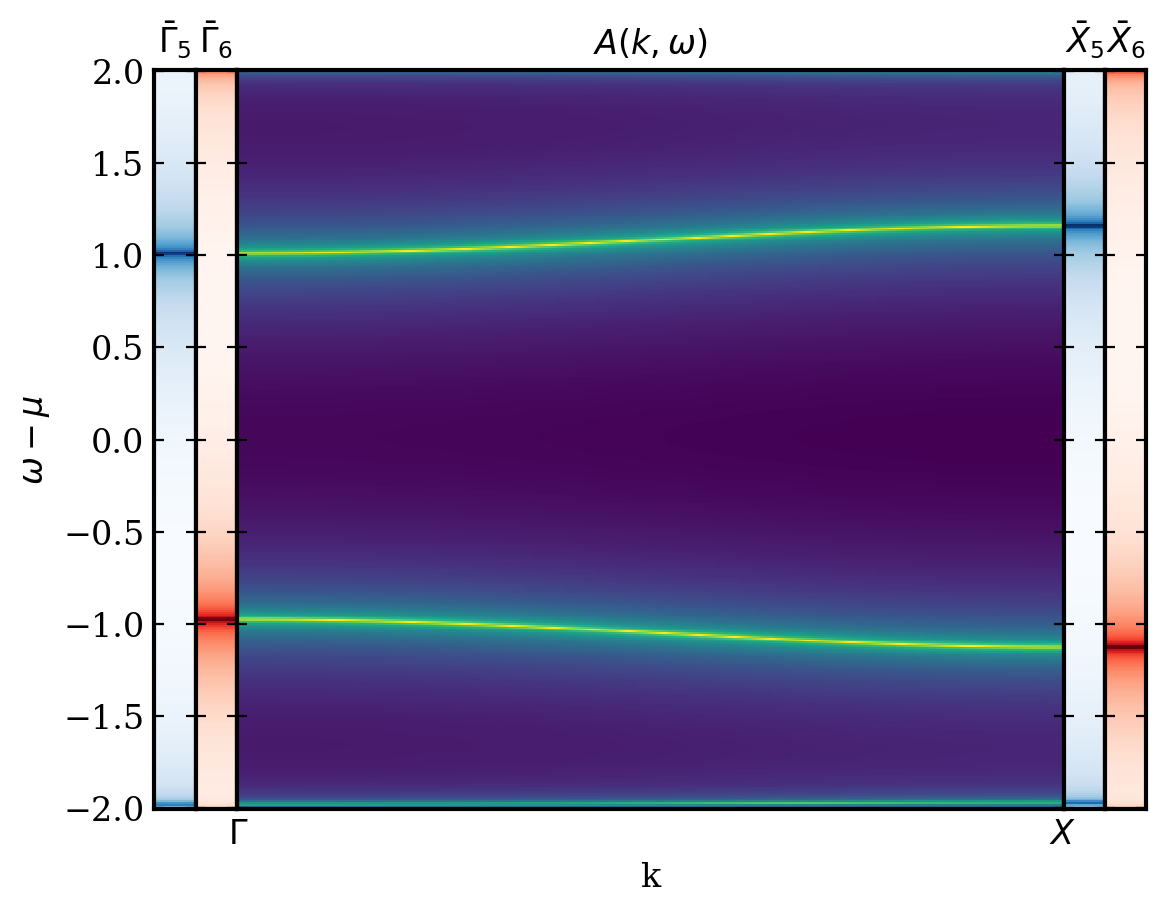}
        \caption{SAI+U at $\phi=\pi/4$.}
    \end{subfigure}\par
    \begin{subfigure}[t]{0.45\linewidth}
        \centering
        \includegraphics[width=\linewidth]{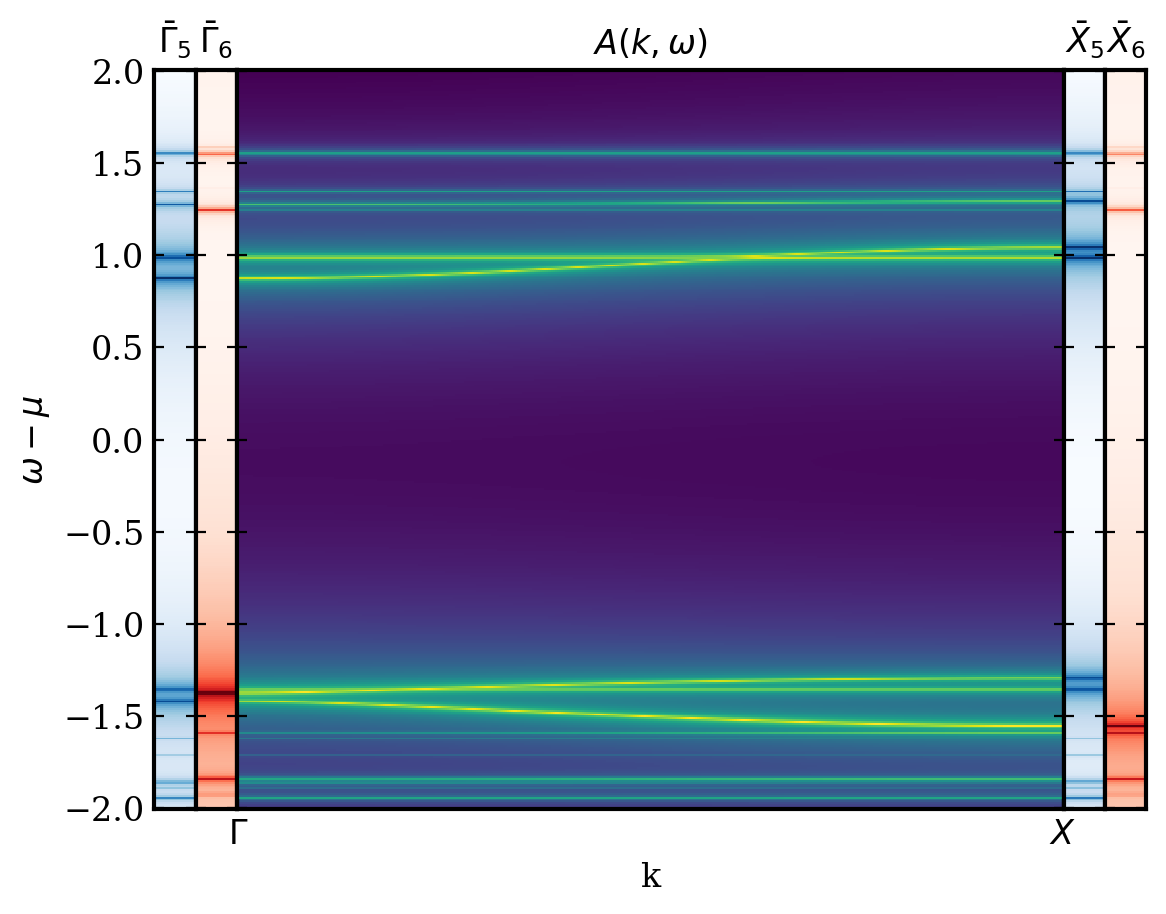}
        \caption{Mott-II at $\phi=(1/2- 10^{-4})\pi$.}
        \label{fig:MottII_spectral}
    \end{subfigure}
    \caption{Representative spectral functions for each phase at $U=4$. The side-panels represent the spectral weight at the high symmetry points separated by irrep of the little group.}
    \label{fig:spectrals}
\end{figure}

Having established the many-body phase diagram, we now examine its single-particle manifestation through the spectral function. Representative spectral functions are presented for each phase of interest in figure \ref{fig:spectrals}. The spectral weight labelled by irrep is represented in the side-panels of figure~\ref{fig:spectrals}. As can be expected for an ED CDMFT simulation of an interacting system, the spectral function possesses many poles approaching the smearing of spectral weight that is observed experimentally in ARPES. 

As anticipated from the above considerations, in the Mott phases the spectral weight of the parent metallic bands is redistributed into upper and lower Hubbard bands, inheriting the symmetry character of the parent metallic phases. Conversely, since the SAI+U phase is adiabatically connected to a non-interacting insulator, the spectral weight has sharp quasiparticle-like peaks.

Both Mott-I(II) phases are related (on the level of their non-interacting parameter sets) to the metallic phases Metal-I(II) (Appendix \ref{annex:non_interacting_spectral}). The non-interacting spectral weight forming the half-filled metallic bands are of a single spatial parity at high symmetry points in both metals (cf. figures \ref{fig:MetalI_spectral} and \ref{fig:MetalII_spectral}). After undergoing the Mott metal-insulator transition \cite{NFMott}, we expect the spectral weight from these metallic bands to be split across the single particle gap into upper and lower Hubbard bands and thus for the irrep to be spread apart as can be seen in figures \ref{fig:MottI_spectral} and \ref{fig:MottII_spectral}.

It is interesting to note that the SAI+U still has clean and seemingly sharp spectral weight near the chemical potential as in the non-interacting SAI phase (figure \ref{fig:SAI_spectral}, Appendix \ref{annex:non_interacting_spectral}). This is consistent with the fact that the SAI+U phase is not a Mott insulator but is rather adiabatically connected to the non-interacting SAI phase \cite{MikelPhD}. The subtle asymmetry of the spectral function about $\omega - \mu = 0$ can be interpreted as an artefact stemming from the finite bath used in ED. In particular, this symmetry is not strictly enforced in the calculations allowing for slight breaking in the fitting procedure.

\subsection{Effective one-body orbitals}\label{subsec:results_one_body_orbitals}

\begin{figure}[tb!]
    \centering
    \begin{subfigure}[t]{0.45\linewidth}
        \centering
        \begin{subfigure}[t]{0.49\linewidth}
            \centering
            \includegraphics[width=\linewidth]{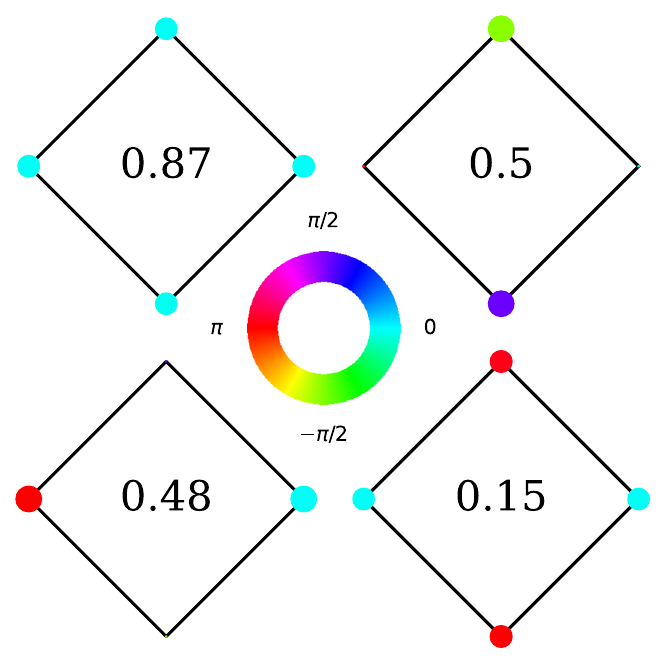}
            \caption*{$\sigma=\,\uparrow$}
        \end{subfigure}%
        ~
        \begin{subfigure}[t]{0.49\linewidth}
            \centering
            \includegraphics[width=\linewidth]{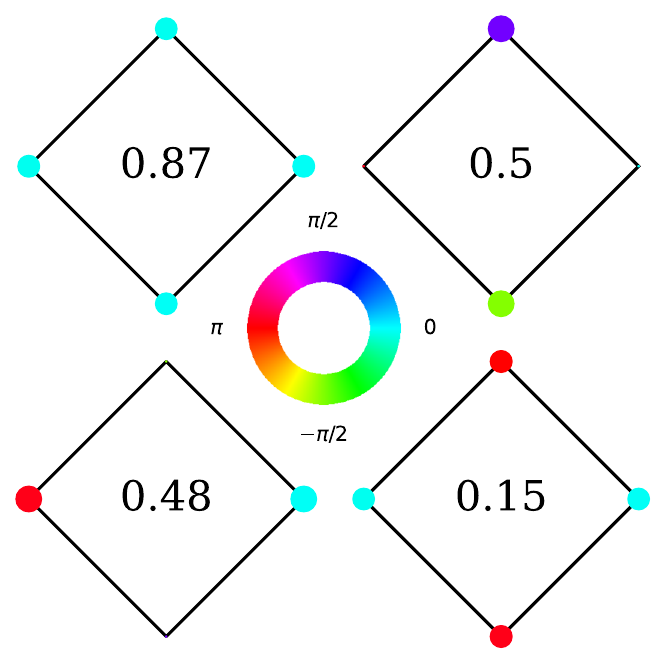}
            \caption*{$\sigma=\,\downarrow$}
        \end{subfigure}
        \caption{Mott-I at $\phi=10^{-4}\pi$.}
        \label{fig:MottI_orbitals}
    \end{subfigure}
    \begin{subfigure}[t]{0.45\linewidth}
        \centering
        \begin{subfigure}[t]{0.49\linewidth}
            \centering
            \includegraphics[width=\linewidth]{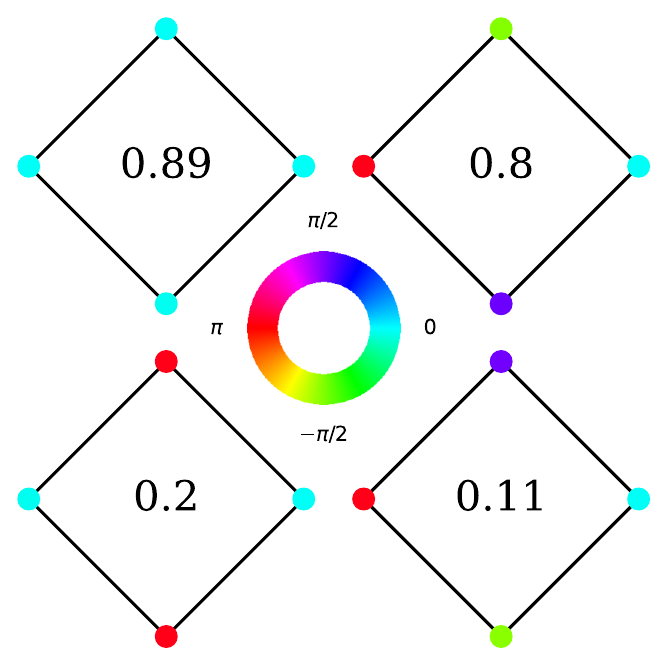}
            \caption*{$\sigma=\,\uparrow$}
        \end{subfigure}%
        ~
        \begin{subfigure}[t]{0.49\linewidth}
            \centering
            \includegraphics[width=\linewidth]{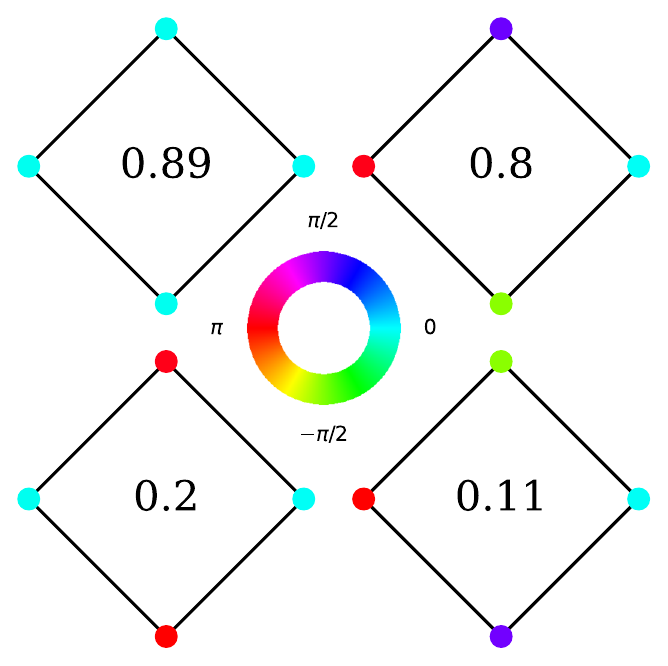}
            \caption*{$\sigma=\,\downarrow$}
        \end{subfigure}
        \caption{SAI+U at $\phi=\pi/4$.}
        \label{fig:SAIU_orbitals}
    \end{subfigure}\par
        \begin{subfigure}[t]{0.45\linewidth}
        \centering
        \begin{subfigure}[t]{0.49\linewidth}
            \centering
            \includegraphics[width=\linewidth]{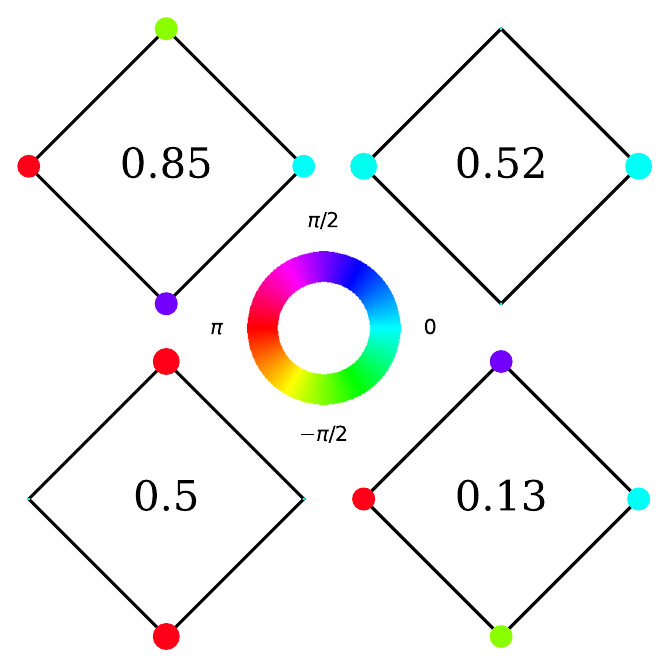}
            \caption*{$\sigma=\,\uparrow$}
        \end{subfigure}%
        ~
        \begin{subfigure}[t]{0.49\linewidth}
            \centering
            \includegraphics[width=\linewidth]{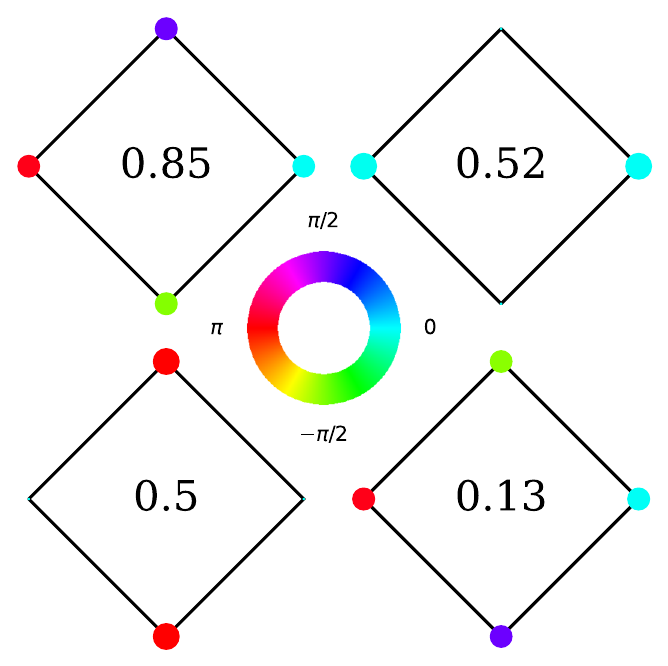}
            \caption*{$\sigma=\,\downarrow$}
        \end{subfigure}
        \caption{Mott-II at $\phi=(1/2- 10^{-4})\pi$.}
        \label{fig:MottII_orbitals}
    \end{subfigure}
    \caption{Effective one-body orbitals of the occupied states for all phases at $U=4$. Classical probability is indicated in the center of each orbital. Amplitude and phase of the orbital components associated with each site are represented by the radius and the colour of the circle at each orbital site.}
    \label{fig:orbitals}
\end{figure}

The effective one-body orbitals provide a complementary perspective in the orbital basis. Representative effective orbitals for each phase are shown in Fig.~\ref{fig:orbitals}. Their spatial character enables a qualitative distinction between the three phases.

Starting with Mott-I (fig. \ref{fig:MottI_orbitals}), one can identify (in order) effective orbitals akin to $s$, $p_x$, $p_y$, $d_{x^2-y^2}$. Effective orbitals for both spins are entirely equivalent since the graph is taken at $\phi\simeq0$. As $\phi$ is increased into the SAI+U phase (fig. \ref{fig:SAIU_orbitals}), the $s$ and $d_{x^2-y^y}$ type orbitals are preserved while the $p_x$ and $p_y$ orbitals of Mott-I are converted into clockwise or counterclockwise rotating modes. Which one is favoured in the statistics depends on the spin sector since these are driven by SOC. At $\phi=\pi/2$, the system is in Mott-II (fig. \ref{fig:MottII_orbitals}). Although the rotating orbitals are comparable to the SAI+U phase, the $s$ and $d_{x^2-y^2}$ are instead replaced by spatially even pseudo-p orbitals.

\subsection{Density matrix purity}\label{subsec:results_purity}

A simple scalar diagnostic of the phase transitions is provided by the purity of the 1RDM. Because the 1RDM is generally a mixed state (Section \ref{subsec:onerdm}), its purity measures the deviation of the many-body ground state from a single Slater determinant (Section \ref{subsubsec:purity}).

\begin{figure}[h!]
    \centering
    \includegraphics[width=0.5\linewidth]{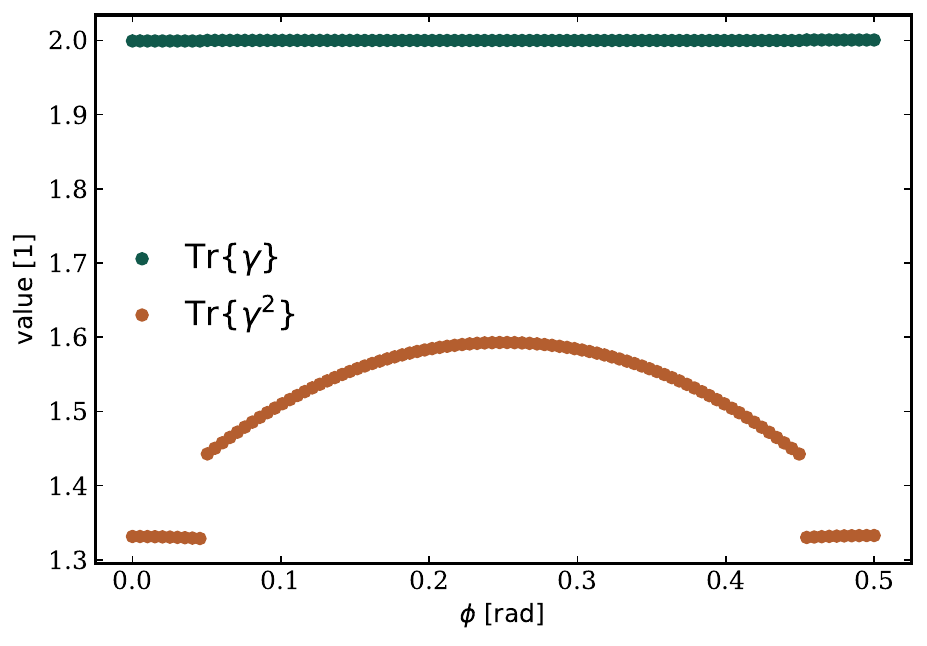}
    \caption{Average of the traces and purities of the orbital 1RDM at $\Gamma$ and $X$ as a function of SOC strength $\phi$.}
    \label{fig:purity_sweep}
\end{figure}

In figure \ref{fig:purity_sweep}, the trace and the purity of the orbital 1RDM are plotted as a function of SOC strength. The trace is constant across values of $\phi$, which is to be expected since the filling of the model is given by the trace of the orbital 1RDM divided by the number of orbitals in the unit cell. The purity is discontinuous at two points in the graph corresponding to the transition between Mott and SAI. Moreover, it is symmetric around $\phi=\pi/4$, which is to be expected under examination of \eqref{eq:hopping_matrix}. These discontinuities occur at values of $\phi$ consistent with the phase boundaries established by DMRG in Section~\ref{subsec:results_TN}, providing an independent single-particle confirmation of the phase transitions.

\section{Conclusion} \label{conclusions} 

In this work, we have applied the interacting one-body picture to the Hubbard diamond chain, a model hosting three distinct correlated phases driven by the interplay between Hubbard interactions and spin-orbit coupling. Using DMRG, we first confirmed that the three-phase structure reported in \cite{MikelPhD} persists at $t_2/|t_1| = 0.8$, with phase boundaries shifting toward smaller SOC strengths as $t_2$ increases.

The three methods of the interacting one-body picture provide a complementary diagnostic of the interacting phases. First, symmetry-resolved spectral functions cleanly distinguish the Mott phases from the SAI+U through the irrep structure of the spectral function. Second, the effective one-body orbitals yield an orbital basis characterization, providing qualitative discrimination between all three phases. Finally, the sharpest diagnostic, the purity of the 1RDM becomes discontinuous precisely at the Mott-SAI+U boundaries. It is detectable purely at the single-particle level without requiring the many-body gap or mirror expectation values.

These results suggest that the interacting one-body picture can serve as a practical bridge between ab initio Green's function and experiment, since symmetry-resolved spectral functions are directly comparable to ARPES data and the purity is straightforwardly computable from any Green's function workflow. Extending these tools to realistic materials represents a natural next step. In this sense, the interacting one-body picture provides a practical and transferable framework for analyzing correlated insulating phases from single-particle quantities.

\section*{Acknowledgements}
We wish to thank David Sénéchal and André-Marie Tremblay for their time and many crucial discussions in the beginning of this work. We also wish to thank Juan Luis Mañes for his expedient help in analyzing the symmetry of the model. Moreover, we thank Johannes Hauschild for his help with the TenPy library. Finally, we recognize Antoine de Lagrave's and Jérôme Leblanc's efficient help with the technical details of CDMFT and integration contours respectively.


\paragraph{Funding information}
T.N.D. acknowledges the support of the Natural Sciences and Engineering Research Council of Canada (NSERC), the Fonds de recherche du Québec - Nature et technologies (FRQNT) and the Fondation de l'Université de Sherbrooke (FUS).

M.G.V. acknowledges the support of PID2022-142008NB-I00 funded by MICIU/AEI/10.130\allowbreak39/501100011033 and FEDER, UE, the Canada Excellence Research Chairs Program for Topological Quantum Matter and to Diputacion Foral de Gipuzkoa Programa Mujeres y Ciencia.

This work has been financially supported by the Ministry for Digital Transformation and of Civil Service of the Spanish Government through the QUANTUM ENIA project call - Quantum Spain project, and by the European Union through the Recovery, Transformation and Resilience Plan - NextGenerationEU within the framework of the Digital Spain 2026 Agenda.

\begin{appendix}
\numberwithin{equation}{section}


\section{Finite temperature complex frequency Green's function in Lehmann's representation}\label{annex:SPGF_derivation}

We start from the definition of the spectral function in \cite{dionne_pyqcm_2023}
\begin{align}
    A_{\mu\nu}(t) = \left\langle\left\{\mathrm{c}_\mu(t),\mathrm{c}^\dagger_\nu(0)\right\}\right\rangle
\end{align}
Focusing on the electron contribution $A_{\mu\nu}^{(e)}(t)$, we can see that at thermal equilibrium for a time independent hamiltonian
\begin{align}
    A_{\mu\nu}^{(e)}(t) &= \mathcal{Z}^{-1}\operatorname{Tr}_{\mathcal{H}_N}\acc{\Exp{-\beta\op{H}}\Exp{i\op{H}t}c_\mu\Exp{-i\op{H}t} c_\nu^\dagger}\notag\\
    &= \mathcal{Z}^{-1}\sum_{n}\bra{n^{(N)}}\Exp{-\beta\op{H}}\Exp{i\op{H}t}c_\mu\Exp{-i\op{H}t} c_\nu^\dagger\ket{n^{(N)}}\notag\\
    &= \mathcal{Z}^{-1}\sum_{n,m}\Exp{-\beta E_n^{(N)
    }}\Exp{i(E_n^{(N)} - E_m^{(N+1)})t} \bra{n^{(N)}}c_\mu\ket{m^{(N+1)}}\!\!\bra{m^{(N+1)}}c_\nu^\dagger\ket{n^{(N)}}
\end{align}
where $\mathcal{H}^{(N)}$ is the subspace of $N$-particle states. The spectral function in frequency space is given by
\begin{align}
    A_{\mu\nu}^{(e)}(\omega) &= \mathcal{Z}^{-1}\int_{-\infty}^{\infty}\D{t}\sum_{n,m}\Exp{-\beta E_n^{(N)
    }}\Exp{i(\omega + E_n^{(N)} - E_m^{(N+1)})t}\bra{n^{(N)}}c_\mu\ket{m^{(N+1)}}\!\!\bra{m^{(N+1)}}c_\nu^\dagger\ket{n^{(N)}}\notag\\
    &= \mathcal{Z}^{-1}\sum_{n,m}\Exp{-\beta E_n^{(N)
    }}2\pi\delta\p{\omega - E_n^{(N)} + E_m^{(N+1)}}\bra{n^{(N)}}c_\mu\ket{m^{(N+1)}}\!\!\bra{m^{(N+1)}}c_\nu^\dagger\ket{n^{(N)}}
\end{align}
The electronic contribution to the Green's function is then obtained \cite{dionne_pyqcm_2023}
\begin{align}
    G_{\mu\nu}^{(e)}(z) &= \int_{-\infty}^{\infty}\frac{\D{\omega}}{2\pi}\frac{A_{\mu\nu}^{(e)}(\omega)}{z - \omega}\notag\\
    &= \mathcal{Z}^{-1}\sum_{n,m}\Exp{-\beta E_n^{(N)
    }}\frac{\bra{n^{(N)}}c_\mu\ket{m^{(N+1)}}\!\!\bra{m^{(N+1)}}c_\nu^\dagger\ket{n^{(N)}}}{z - E_m^{(N+1)} + E_n^{(N)}}\label{eq:Lehmann_form}
\end{align}
An analogous procedure can be performed on the hole contribution then resulting in the full complex-frequency single-particle Green's function:
\begin{align}
    G_{\mu\nu}(z) &= \mathcal{Z}^{-1}\sum_{n}\Exp{-\beta E_n^{(N)
    }}\Bigg[\sum_{m}\frac{\bra{n^{(N)}}c_\mu\ket{m^{(N+1)}}\!\!\bra{m^{(N+1)}}c_\nu^\dagger\ket{n^{(N)}}}{z - E_m^{(N+1)} + E_n^{(N)}}\notag\\
    &+ \sum_{\ell}\frac{\bra{n^{(N)}}c_\nu^\dagger\ket{\ell^{(N-1)}}\!\!\bra{\ell^{(N-1)}}c_\mu\ket{n^{(N)}}}{z - E_n^{(N)} + E_\ell^{(N-1)}}\Bigg]\label{eq:full_Lehmann}
\end{align}
Note that the complex frequency Green's function is a meromorphic complex function containing only simple poles on the real axis \cite{gurarie_single-particle_2011} which is clear from its above form. Moreover, It can be shown that the complex frequency Green's function is the only one possessing the correct properties: giving the correct Green's functions (retarded, advanced, Matsubara, \textit{etc.}) and simultaneously obeying the appropriate boundary conditions \cite{rickayzen_greens_2013, 10.1063/1.1703704}.


\section{Symmetry of the general complex frequency SPGF} \label{annex:symmetry_SPGF}

In this section, we show that the single particle Green's function transforms like a fermionic bilinear under unitary symmetries. This derivation, in the interest of generality, is carried out for arbitrary frequency and temperature. \textbf{\textit{Note that this proof can be trivially extended to n-particle Green's functions by grouping indices.}}

Given the representation of the unitary transform on the ladder operators \eqref{eq:ladders_under_unitaries}, one can naturally define that the hamiltonian is symmetric under the transformation if it commutes with the representation of the entire group.

However, it is also well known \cite{coleman_introduction_2015} that a system of interacting particles can develop a ground state with lower symmetry than the hamiltonian. For this purpose, we also require that the symmetry preserves the subspace of Hilbert space containing the ground state with spontaneously broken symmetry. For example, if the system spontaneously develops magnetization along an axis, any rotation \textit{around} that axis is preserved. Conversely, the other rotations are no longer generally a symmetry of the Green's function.

Thus, we define \textit{the system} to be symmetric under a unitary transformation as long as:
\begin{align}
    \left[\hat{\mathrm{H}}, \hat{\mathrm{U}}(g)\right] = 0~~\text{and}~~ \hat{\mathrm{U}}:\mathcal{H}_{\text{SSB}}\to\mathcal{H}_{\text{SSB}} ~\forall~g\in G
\end{align}
Now, we can examine the transformation of the Green's function under the ladder operator representation. Starting from the electronic contribution (see Appendix \ref{annex:SPGF_derivation}) of the LHS in \eqref{eq:sym_garochée}, we can use the Lehmann form \eqref{eq:Lehmann_form} to obtain:
\begin{align}
    \sum_{\alpha\beta}U_{\mu\alpha}G^{(e)}_{\alpha\beta}(z)U_{\beta\nu}^*
    &=\mathcal{Z}^{-1}\sum_{n m}\Exp{-\beta E_n^{(N)
    }}\frac{\bra{n^{(N)}}\sum_\alpha U_{\mu\alpha}c_\alpha\ket{m^{(N+1)}}\!\!\bra{m^{(N+1)}}\sum_\beta c_\beta^\dagger U_{\beta\nu}^*\ket{n^{(N)}}}{z - E_m^{(N+1)} + E_n^{(N)}}\notag\\
    &=\mathcal{Z}^{-1}\sum_{n m}\Exp{-\beta E_n^{(N)
    }}\frac{\bra{n^{(N)}}\hat{\mathrm{U}}c_\mu\hat{\mathrm{U}}^\dagger\ket{m^{(N+1)}}\!\!\bra{m^{(N+1)}}\hat{\mathrm{U}}c_\beta^\dagger \hat{\mathrm{U}}^\dagger\ket{n^{(N)}}}{z - E_m^{(N+1)} + E_n^{(N)}}\label{eq:incomplete_sentence}
\end{align}

Since the unitary transformation commutes with both the hamiltonian and the number operator, the action of the $\hat{\mathrm{U}}$ on the set $\{\ket{n^{(N)}}\}$ forms a unitary representation that is closed on the subspace of degenerate states (energy $E_n^{(N)}$) and the subspace of $N$ particles. Precisely,
\begin{align}
    \hat{\mathrm{U}}\ket{n^{(N)},i} &= \sum_{j\in\text{deg}(n)}\mathcal{U}_{ij}^{(N)}(n)\ket{n^{(N)},j} & \left[\mathcal{U}\right]^\dagger\!\left[\mathcal{U}\right]&=\mathbf{1}_{\text{deg}(n)}
\end{align}
We can then show that \eqref{eq:incomplete_sentence} is indeed invariant. To see this, the sums over the fixed particle number spectrum are replaced by a sum over energies and a sum over degeneracies:
\begin{align}
    &\sum_{\substack{n\in\mathcal{H}^{(N)}|_{E_n}\\m\in\mathcal{H}^{(N+1)}|_{E_m}}}\,\sum_{\substack{i\in\text{deg(n)}\\j\in\text{deg(m)}}}\Exp{-\beta E_n^{(N)
    }}\frac{\bra{n^{(N)},i}\hat{\mathrm{U}}c_\mu\hat{\mathrm{U}}^\dagger\ket{m^{(N+1)},j}\!\!\bra{m^{(N+1)},j}\hat{\mathrm{U}}c_\beta^\dagger \hat{\mathrm{U}}^\dagger\ket{n^{(N)},i}}{z - E_m^{(N+1)} + E_n^{(N)}}\notag\\
    &=\sum_{\substack{n\in\mathcal{H}^{(N)}|_{E_n}\\m\in\mathcal{H}^{(N+1)}|_{E_m}}}\,\sum_{\substack{i,a,b\in\text{deg(n)}\\j,c,d\in\text{deg(m)}}}\Exp{-\beta E_n^{(N)
    }}\frac{\bra{n^{(N)},a}\mathcal{U}_{ia}^{(N)}(n)c_\mu\mathcal{U}_{cj}^{(N+1)*}(m)\ket{m^{(N+1)},c}}{z - E_m^{(N+1)} + E_n^{(N)}}\notag\\
    &\times \bra{m^{(N+1)},d}\mathcal{U}_{jd}^{(N+1)}(m)c_\beta^\dagger \mathcal{U}_{bi}^{(N)*}(n)\ket{n^{(N)},b}\notag\\
    &=\sum_{\substack{n\in\mathcal{H}^{(N)}|_{E_n}\\m\in\mathcal{H}^{(N+1)}|_{E_m}}}\,\sum_{\substack{i,a,b\in\text{deg(n)}\\j,c,d\in\text{deg(m)}}}\Exp{-\beta E_n^{(N)
    }}\mathcal{U}_{cj}^{(N+1)*}\mathcal{U}_{jd}^{(N+1)}\mathcal{U}_{bi}^{(N)*}\mathcal{U}_{ia}^{(N)}\frac{\bra{n^{(N)},a}c_\mu\ket{m^{(N+1)},c}}{z - E_m^{(N+1)} + E_n^{(N)}}\notag\\
    &\times \bra{m^{(N+1)},d}c_\beta^\dagger\ket{n^{(N)},b}\notag\\
    &=\sum_{\substack{n\in\mathcal{H}^{(N)}|_{E_n}\\m\in\mathcal{H}^{(N+1)}|_{E_m}}}\,\sum_{\substack{i,a,b\in\text{deg(n)}\\j,c,d\in\text{deg(m)}}}\Exp{-\beta E_n^{(N)
    }}\delta_{cd}\delta_{ba}\frac{\bra{n^{(N)},a}c_\mu\ket{m^{(N+1)},c}\!\!\bra{m^{(N+1)},d}c_\beta^\dagger\ket{n^{(N)},b}}{z - E_m^{(N+1)} + E_n^{(N)}} = G^{(e)}_{\mu\nu}(z)
\end{align}
An identical derivation holds for the other term in \eqref{eq:full_Lehmann} proving the statement in this section. The restriction imposed to the unitary symmetry demanding closure over the subspace of allowed states by the possibly spontaneously broken symmetry of the system is crucial in making the above step.


\section{Symmetry analysis of the Hubbard Diamond Chain}\label{annex:sym_analysis}

Here, the symmetry-allowed couplings for the Hubbard Diamond Chain are derived. We demand that the model be symmetric under space group Pmmm + time reversal. Knowing how fermions transform under reflections and time reversal, we can lay out the forms of the relevant operators acting in spin-space:
\begin{align}
    M_x &= \begin{pmatrix}
        0&-i\\-i&0
    \end{pmatrix} &
    M_y &= \begin{pmatrix}
        0&-1\\1&0
    \end{pmatrix} &
    M_z &= \begin{pmatrix}
        -i&0\\0&i
    \end{pmatrix} &
    \mathcal{T} &= \begin{pmatrix}
        0&1\\-1&0
    \end{pmatrix}\mathcal{K}
\end{align}
where the basis is taken to be the spin-1/2 projections along $z$ and $\mathcal{K}$ is the complex conjugation operator. The spatial action of the operators on the positions of the model's orbitals acts intuitively. For example, the $M_x$ operator inverts the positions of the unit cells along $x$ and maps orbitals $1\leftrightarrow 3$, leaving 2 and 4 invariant (cf.\ figure~\ref{fig:HDC_chain}). First and foremost, given the orientation of the model running along $x$ in the $xy$ plane, all couplings must be invariant under $M_z$. So starting with a general hopping matrix along any given link expressed in the spin basis (where $a,b,c,d\in\mathbb{C}$):

\begin{align}
    \tilde{t} = M_z\tilde{t}M_z^{-1} = \begin{pmatrix}
        -i&0\\0&i
    \end{pmatrix}
    \begin{pmatrix}
        a&b\\c&d
    \end{pmatrix}
    \begin{pmatrix}
        i&0\\0&-i
    \end{pmatrix}
    = \begin{pmatrix}
        a&-b\\-c&d
    \end{pmatrix}~\Rightarrow~\tilde{t} = \begin{pmatrix}
        a&0\\0&d
    \end{pmatrix}
\end{align}
Now, time reversal symmetry can be used to further constrain the form of the couplings:
\begin{align}
    \tilde{t} = \mathcal{T}\tilde{t}\mathcal{T}^{-1} = \begin{pmatrix}
        0&1\\-1&0
    \end{pmatrix}
    \begin{pmatrix}
        a^*&0\\0&d^*
    \end{pmatrix}
    \begin{pmatrix}
        0&-1\\1&0
    \end{pmatrix}
    =\begin{pmatrix}
        d^*&0\\0&a^*
    \end{pmatrix}~\Rightarrow~\tilde{t} = \begin{pmatrix}
        a&0\\0&a^*
    \end{pmatrix}
\end{align}
At this stage, only one complex parameter remains for any possible coupling in the plane.

The purely vertical and horizontal links in the model are now shown to be real:
\begin{align}
    \tilde{t}_{\text{vert}} &= M_x\tilde{t}_{\text{vert}}M_x^{-1} ~\Rightarrow~ \tilde{t}_{\text{vert}} = \begin{pmatrix}
        a&0\\0&a
    \end{pmatrix}~,~a\in\mathbb{R} & \tilde{t}_{\text{hor}} &= M_y\tilde{t}_{\text{hor}}M_y^{-1} ~\Rightarrow~ \tilde{t}_{\text{hor}} = \begin{pmatrix}
        b&0\\0&b
    \end{pmatrix}~,~b\in\mathbb{R}
\end{align}
It is to be noted that the next-neighbour intra-diamond couplings have no need to be identical, we have simply chosen to do so out of simplicity in this work.

The diagonal links do not share this same symmetry and must be examined slightly more carefully. Consider the diagonal link that takes site 1 to site 2 ($\tilde{t}_{12}$). Upon mirror reflection $M_x$, the coupling is mapped to $\tilde{t}_{32}^*$. However, since $\tilde{t}_{32}^*=\tilde{t}_{23}$, we can establish that $\tilde{t}_{12}=\tilde{t}_{23}$. Through the use of similar arguments, it can be verified that $\tilde{t}_{12}=\tilde{t}_{23}=\tilde{t}_{34}=\tilde{t}_{41}$.

In summary, all couplings except for the diagonal ones are \textit{real} and are identical no matter the spin projections:
\begin{align}
    \tilde t_2\sim \tilde t_3\sim \mathbbm{1}
\end{align}
Moreover, the diagonal links are all identical with respect to a given order around the sites of the diamond and specified by a single complex number:
\begin{align}
    \tilde t_1 = \begin{pmatrix}
        t_1 & 0 \\0 & t_1^*
    \end{pmatrix}
\end{align}

\section{TeNPy simulation details }\label{annex:TN_details}
This section describes how to reproduce the calculations that led to the results presented in figure \ref{fig:TN_unified}.

\begin{figure}[htb!]
    \centering
    \begin{subfigure}[b]{0.7\linewidth}
        \centering
        \includegraphics[width=\linewidth]{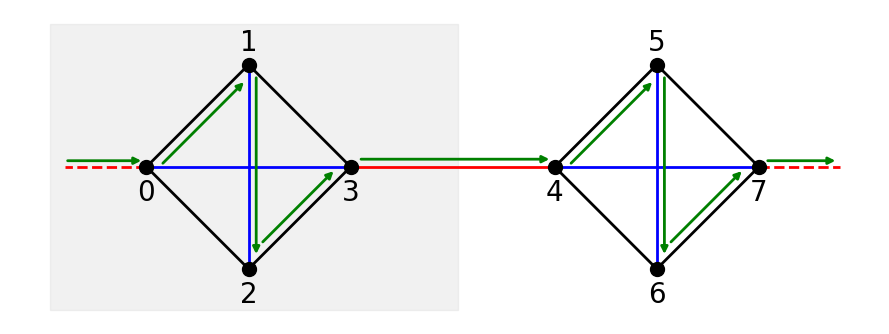}
        \caption{Representation of two unit cells in the Hubbard diamond chain.}
        \label{fig:ordering}
    \end{subfigure}%
    \\
    \begin{subfigure}[b]{0.7\linewidth}
        \centering
        \includegraphics[width=\linewidth]{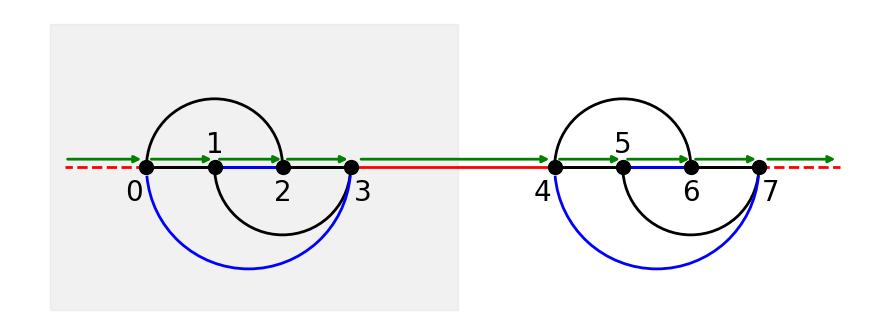}
        \caption{Conversion of the previously shown diamond chain into a one-dimensional chain according to the ordering indicated by the green arrows.}
        \label{fig:ordering_chain}
    \end{subfigure}
    \caption{Ordering of the Hubbard diamond lattice for a chain consisting of two diamonds, and its mapping onto a one-dimensional chain representation required for the application of the DMRG algorithm.}
    \label{fig:ordering_1d}
\end{figure}

The first essential step is to define the lattice structure of the problem. In this case, the diamond lattice is not predefined in TenPy, so it must be constructed manually using the generic \textit{Lattice} class. The next step consists of specifying an ordering of the lattice sites in order to map the two-dimensional structure onto a one-dimensional chain, as required by the DMRG algorithm. This ordering, along with a schematic representation of the mapping from the diamond lattice to its corresponding one-dimensional chain, is shown in Figures~\ref{fig:ordering} and~\ref{fig:ordering_chain}.

Once an appropriate site ordering is defined, we specify the physical properties of each site. For this purpose, we employ the \textit{SpinHalfFermionSite} class, which assigns spin $1/2$ fermionic degrees of freedom to each site and provides the corresponding operators (creation, annihilation, $S_x$,$S_y$,$Sz$, etc.). In addition, it allows us to define conserved quantities such as the total particle number. Enforcing these conservation laws significantly reduces the size of the accessible Hilbert space, improving the computational efficiency of the simulation.

Once the full model is properly defined and the required classes are initialized, we create a half-filled ansatz wavefunction $|\psi_{ansatz}\rangle = | \uparrow, \uparrow, \downarrow, \downarrow \rangle $. With the Hamiltonian already constructed, we employ the \textit{TwoSiteDMRGEngine} to perform the optimization under open boundary conditions.

Several key parameters determine the performance and accuracy of the DMRG engine, and here we highlight the most relevant ones. First, a mixer is introduced at the beginning of the simulation to help avoid convergence to local minima. Second, the diagonalization method employed is the Lanczos algorithm, which efficiently computes the lowest-energy eigenstates.
We also allow the bond dimension to reach values of up to approximately $300$; however, such large values are generally unnecessary due to the use of open boundary conditions, which naturally reduces entanglement near the edges. Finally, the convergence criteria are set to a maximum energy error of $\Delta E= 10^{-8}$ and an entropy change threshold of $\Delta S= 10^{-5}$.

\begin{figure}[htb!]
    \centering
    \includegraphics[width=0.6\linewidth]{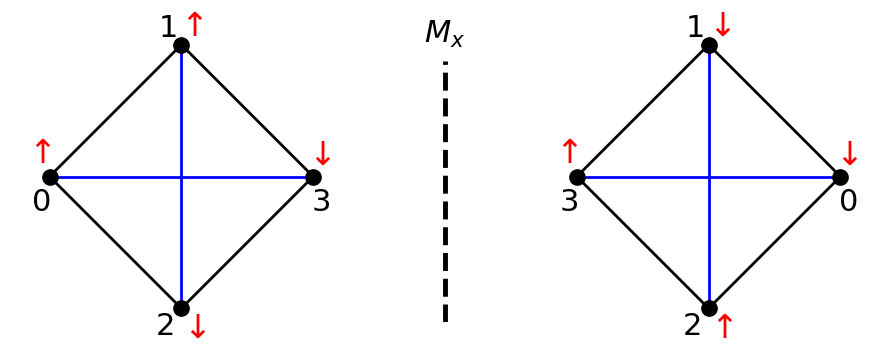}
    \caption{Schematic representation of the mirror operator. In the left part we can see the original configuration and in the right side we can observe the result of applying a mirror operator.}
    \label{fig:Mx_symmetry}
\end{figure}

As explained in Section~\ref{subsec:methods_TN_tenpy}, once a reliable approximation to the ground-state wavefunction has been obtained, we perform a series of operations on the MPS to compute the expectation value of the mirror operator. Since this operator is not predefined in TenPy, it must be constructed from a combination of other operations. First, after creating a copy of the ground-state MPS, we apply a permutation of the lattice sites corresponding to the exchange of site indices between the left and right sides of the system, as illustrated in Figure~\ref{fig:Mx_symmetry}. Next, we apply the $S_x$ operator to flip the spins, which are originally defined along the $z$-direction. The combination of these two operations is equivalent to applying the mirror operator, as shown schematically in figure~\ref{fig:Mx_symmetry}. Finally, by computing the overlap between this transformed state and the original wavefunction, we obtain the expectation value $\langle \tilde{\psi_0}|M_x|\tilde{\psi}_{0}\rangle$.

Afterwards, starting again from the ground-state wavefunction, we consider several possible approaches to compute the first excited state and the corresponding energy gap. In this work, we choose to create a copy of the ground-state MPS and perform a second DMRG calculation using the same configuration as before, except for the inclusion of the additional argument \textit{orthogonal\_to}, which enforces orthogonality with respect to the copied ground-state MPS.

This procedure is repeated for each point in the parameter grid corresponding to figure \ref{fig:TN_unified}. For the plots of the $M_x$ expectation value, calculations were performed for approximately 15.000 independent points, while the energy-gap computations were carried out for 10.000 points. Since the evaluation at each grid point is independent of the others, the entire process can be easily parallelized.


\section{\texorpdfstring{Pyqcm CDMFT details}{Pyqcm CDMFT details}}\label{annex:pyqcm_details}

In section~\ref{annex:pyqcm_impurity}, the details regarding the choice of effective impurity models are given while \ref{annex:pyqcm_params} contains the parameters used in the simulations.

\subsection{\texorpdfstring{Constructing the effective impurity model}{Constructing the effective impurity model}}\label{annex:pyqcm_impurity}\label{subsubsec:methods_cluster_bath_setup}

To apply CDMFT via the use of ED, a choice of cluster, baths and hybridizations must be made \cite{dionne_pyqcm_2023}. Since the phases studied in the model arise from single diamond physics \cite{iraola_towards_2021, soldini_interacting_2024}, it is intuitive to partition the lattice into single diamond clusters. 

\begin{figure}[h!]
    \centering
    \includegraphics[height=0.4\linewidth]{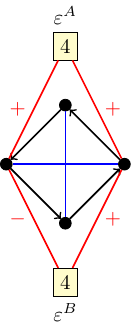}
    \caption{The cluster and bath configuration of the effective AIM solved via ED. 4 sites are used per irrep($A/B$) of C$_2$. The relative signs of the hybridizations are indicated in red adjacent to the respective lines representing the hybridizations.}
    \label{fig:CDMFT_baths}
\end{figure}

Due to the influence of SOC and the lattice, the effective symmetry of the cluster model we consider is C$_2$. The effective AIM was constructed following the irreps of the cluster symmetry group \cite{florezablan2025bathparameterizationmultibandcluster, Koch_2008} as graphically represented in figure~\ref{fig:CDMFT_baths}. In our AIM, four orbitals belong to the diamond cluster, four belong to irrep $A$ in the bath and four more belong to irrep $B$, bringing the impurity size to 12 orbitals. 

\subsection{\texorpdfstring{CDMFT parameters}{CDMFT parameters}}\label{annex:pyqcm_params}

The Matsubara frequency grid is defined by a fictitious inverse temperature (here, $\beta = 50|t_1|^{-1}$), which is used solely to generate a dense grid of imaginary frequencies and does not represent a physical temperature as all calculations are performed at $T=0$. A sharp cutoff is applied at $\omega = \pm 2|t_1|$, beyond which frequencies are excluded from the distance function. The bath parameters are optimized using the BOBYQA algorithm, and self-consistency is considered reached when the bath accuracy falls below $10^{-5}$. Further details on the implementation of these procedures can be found in the pyqcm documentation~\cite{dionne_pyqcm_2023}.


\section{\texorpdfstring{Effects induced by the finite inter-diamond hopping}{Effects induced by the finite inter-diamond hopping}}\label{annex:t3_effects}

\begin{figure}[h]
    \centering
    \includegraphics[height=0.29\linewidth]{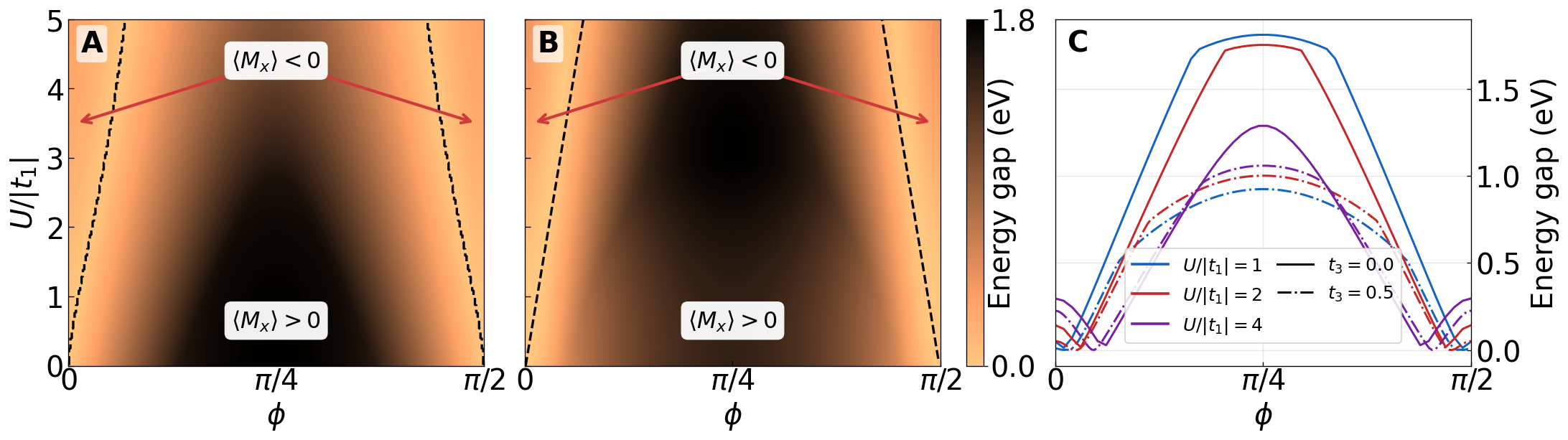}
    \caption[Effect of the inter-diamond hopping on the phase diagram.]{Energy gap between ground state and the first excited state for $t_2/|t_1| = 0.5$, comparing the single-diamond limit ($t_3/|t_1| = 0$, $N = 1$ unit cell, 4 electrons) with a three-diamond chain ($t_3/|t_1| = 0.5$, $N = 3$ unit cells, 12 electrons). \textbf{(A)--(B)} Energy gap as a function of $U/|t_1|$ and $\phi$ for $t_3/|t_1| = 0$ and $t_3/|t_1| = 0.5$, respectively. The dotted lines mark the sign change of $\langle M_x \rangle$, which coincides with the closing of the gap: the central region exhibits $\langle M_x \rangle > 0$, while the lateral regions exhibit $\langle M_x \rangle < 0$. \textbf{(C)} Energy gap as a function of $\phi$ for selected values of $U/|t_1|$ and $t_3/|t_1|$.}
    \label{fig:t3_comparison}
\end{figure}

This section compares the phase diagram obtained in the limit of disconnected diamonds ($t_3=0$, $N=1$ unit cell, 4 electrons) with that of a three-diamond chain ($t_3/|t_1|=0.5$, $N=3$ unit cells, 12 electrons) at fixed $t_2/|t_1|=0.5$. The results are shown in Fig.~\ref{fig:t3_comparison}.

The first thing that we notice is that the three-phase structure, Mott-I ($\phi\approx 0$), SAI+U (intermediate $\phi$) and Mott-II ($\phi\approx\pi/2$), is preserved upon activating $t_3$ (panels (A) and (B) of Fig.~\ref{fig:t3_comparison}). The positions of the phase boundaries, indicated by the dotted lines where $\langle M_x\rangle$ changes sign and the gap closes, remain essentially unchanged, confirming that the finite inter diamond hopping does not alter the critical values of $\phi$ and $U$ at which the transitions occur.

Quantitatively, the finite inter-diamond hopping broadens the single-particle bandwidth, which increases the effective kinetic energy scale of the lattice. This affects the magnitude of the gap without shifting the transition points: the energy gap is uniformly reduced at $t_3/|t_1| = 0.5$ relative to the single-diamond case across all three phases. This is clearly visible in the line cuts of panel (C), where the peaks and troughs shift to lower values in amplitude but the $\phi$ values at which the gap closes are the same for both values of $t_3$. Consistently, the dotted lines in panels (A) and (B) fall at the same values of $\phi$ for both $t_3/|t_1| = 0$ and $t_3/|t_1| = 0.5$, providing direct evidence that the sign change of $\langle M_x\rangle$, and hence the phase transition positions, are unaffected by the inter-diamond coupling. The mirror-symmetry-based classification of the three phases is therefore robust to the inclusion of $t_3$.


\section{One-body reduced density matrix}\label{annex:1RDM}
\begin{figure}[htb!]
    \centering
    \begin{subfigure}[b]{0.45\textwidth}
        \centering
        \includegraphics[width=\textwidth]{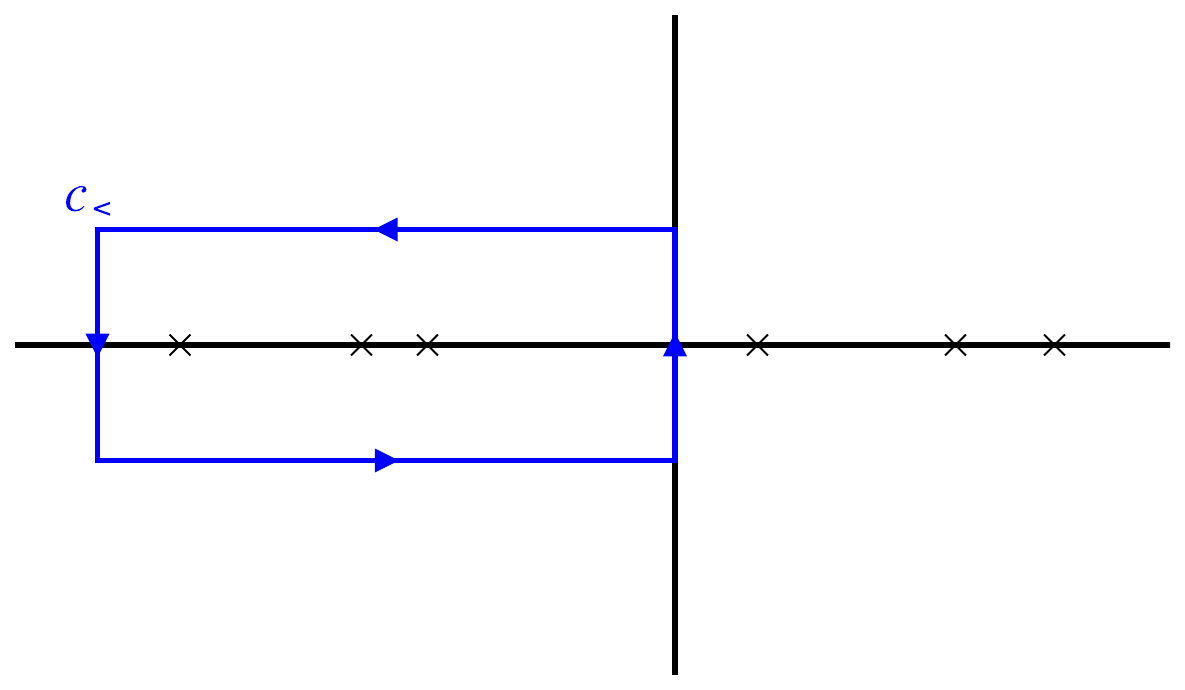}
        \caption{An example contour that can be used to pick up the negative frequency poles of the Green's function.}
        \label{fig:lesser_contour}
    \end{subfigure}
    \hfill
    \begin{subfigure}[b]{0.45\textwidth}
        \centering
        \includegraphics[width=0.5\textwidth]{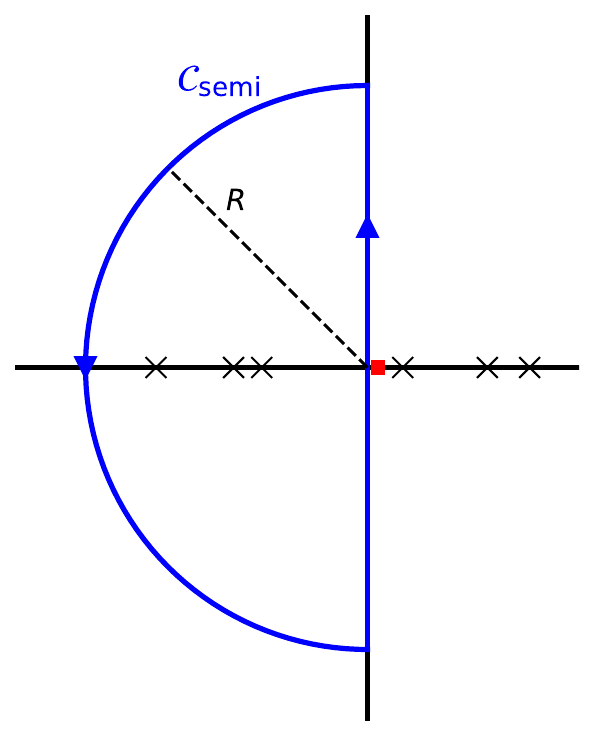}
        \caption{The contour used to obtain an integral over imaginary frequencies. The red pole must lie strictly outside of the semi-circle as to not pick it up in the contour integration.}
        \label{fig:semi_contour}
    \end{subfigure}
    \caption{Contours used in the evaluation of the fermionic bilinear expectation values.}
\end{figure}

Formally, $\hat{\rho}^{(N)}$ is the density operator of the $N$-particle system. The one-body reduced density matrix is obtained by taking the partial trace over the $(N-1)$-particle subspace. This can be shown to reduce to the expectation value of a pair of fermionic ladder operators \cite{gross_many-particle_1991, solovej2014manybody}:
\begin{align}
    \gamma_{\mu\nu} &= \bra{\mu}\operatorname{Tr}_{N-1}\left\{\hat{\rho}^{(N)}\right\}\ket{\nu} \\
    &= \sum_{\alpha_2\dots\alpha_N}\bra{\mu,\alpha_2\dots\alpha_N}\hat{\rho}^{(N)}\ket{\nu,\alpha_2\dots\alpha_N}\notag\\
    &= \sum_{\alpha_2\dots\alpha_N}\bra{\mu,\alpha_2\dots\alpha_N}\left(\sum_{i}p_i\ket{\Psi^{(N)}_i}\!\!\bra{\Psi^{(N)}_i}\right)\ket{\nu,\alpha_2\dots\alpha_N}\notag\\
    &= \sum_{i}p_i\bra{\Psi^{(N)}_i}\hat{\mathrm{c}}_\nu^\dagger\left(\sum_{\alpha_2\dots\alpha_N}\ket{\alpha_2\dots\alpha_N}\!\!\bra{\alpha_2\dots\alpha_N}\right)\hat{\mathrm{c}}_\mu\ket{\Psi^{(N)}_i}\!\!\notag\\
    &= \left\langle\hat{\mathrm{c}}_\nu^\dagger\hat{\mathrm{c}}_\mu\right\rangle\label{eq:partial_trace}
\end{align}
where $\sum_{\alpha_2\dots\alpha_N}\ket{\alpha_2\dots\alpha_N}\!\!\bra{\alpha_2\dots\alpha_N}=\mathbbm{1}\in\mathcal{H}^{(N-1)}$. In practice, one does not generally possess the density operator for the entire system. However, this information can be readily extracted from the single particle Green's function. It is clear from it's Lehmann form \eqref{eq:Lehmann_form} that one can pick up the poles on the negative frequency axis by choosing an appropriate contour ($\mathcal{C}_<$) to obtain the above expectation value \cite{rickayzen_greens_2013, dionne_pyqcm_2023}:
\al{
    \left\langle\hat{\mathrm{c}}_\nu^\dagger\hat{\mathrm{c}}_\mu\right\rangle &= \oint_{\mathcal{C}_<}\frac{\text{d}z}{2\pi i}G_{\mu\nu}(z)
}
The problem can be made numerically tractable by deforming the contour into an infinite semicircle going up the imaginary axis and following an arc downwards at infinity in the half-plane $\mathbb{R}^{-}+i\mathbb{R}$ provided one inserts a simple pole at $p\in\mathbb{R}^+$ to cancel the contribution on the arc (see figure \ref{fig:semi_contour}) \cite{dionne_pyqcm_2023}.

Explicitely, we can see that:
\al{
    \oint_{\mathcal{C}_<}\frac{\text{d}z}{2\pi i}G_{\mu\nu}(z) = \oint_{\mathcal{C}_\text{semi}}\frac{\text{d}z}{2\pi i}\left[G_{\mu\nu}(z)-\frac{\delta_{\mu\nu}}{z - p}\right]
}
Since $\frac{1}{z-p}=\frac{1}{z}+\frac{p}{z^2}+\dots$ around $|z|\to\infty$, this term perfectly cancels out the simple pole that the Green's function has at infinity since it is known to go as $G_{\mu\nu}(z)\to\delta_{\mu\nu}/z$ for large frequencies \cite{rickayzen_greens_2013}. Conveniently, this implies that the contribution on the arc of the large semicircle is exactly zero when its radius is taken to be infinite since:
\al{
    &\int_{\mathcal{C}_\text{arc}}\frac{\text{d}z}{2\pi i}\left[G_{\mu\nu}(z)-\frac{\delta_{\mu\nu}}{z - p}\right]\notag\\
    &= \lim_{R\to\infty}\int\limits_{\pi/2}^{-\pi/2}\frac{R\Exp{i\theta}\text{d}\theta}{2\pi}\left[G_{\mu\nu}(R\Exp{i\theta})-\frac{\delta_{\mu\nu}}{R\Exp{i\theta} - p}\right]\notag\\
    &= \lim_{R\to\infty}\int\limits_{\pi/2}^{-\pi/2}\frac{\text{d}\theta}{2\pi}\left[\delta_{\mu\nu}-\frac{\delta_{\mu\nu}}{1 - p/R\Exp{i\theta}}\right] = 0
}
Hence, all that remains is an integration over the purely imaginary frequencies. Combining this result with \eqref{eq:partial_trace} shows that:
\begin{align}
    \gamma_{\mu\nu} = \int\limits_{-\infty}^{+\infty}\frac{\text{d}\omega}{2\pi}\left[G_{\mu\nu}(i\omega)-\frac{\delta_{\mu\nu}}{i\omega - p}\right]
\end{align}
which is how this is computed in practice from the complex Green's function at $T=0$.


\section{Spectral functions of non-interacting phases}\label{annex:non_interacting_spectral}

\begin{figure}[H]
    \centering
    \begin{subfigure}[b]{0.45\linewidth}
        \centering
        \includegraphics[width=\linewidth]{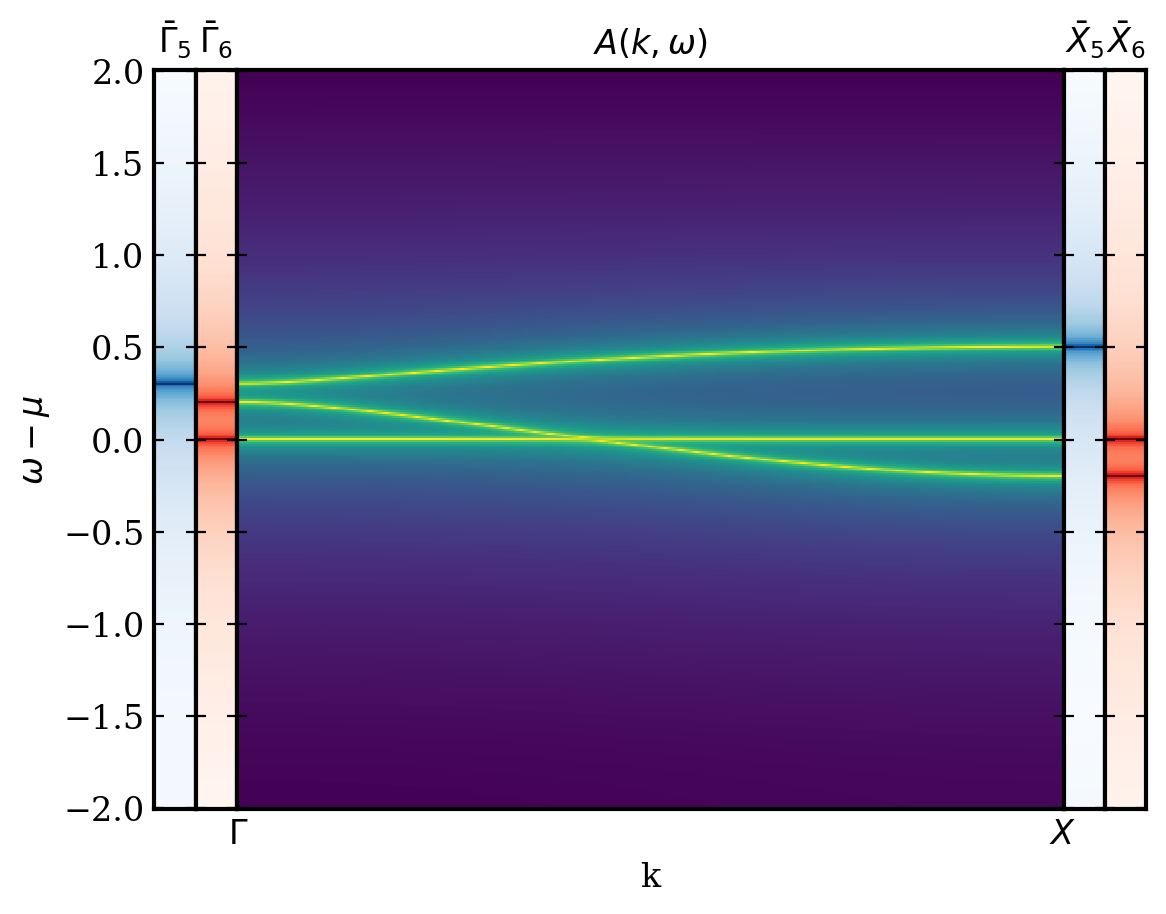}
        \caption{Metal-I $(\phi=10^{-4})$}
        \label{fig:MetalI_spectral}
    \end{subfigure}%
    ~
    \begin{subfigure}[b]{0.45\linewidth}
        \centering
        \includegraphics[width=\linewidth]{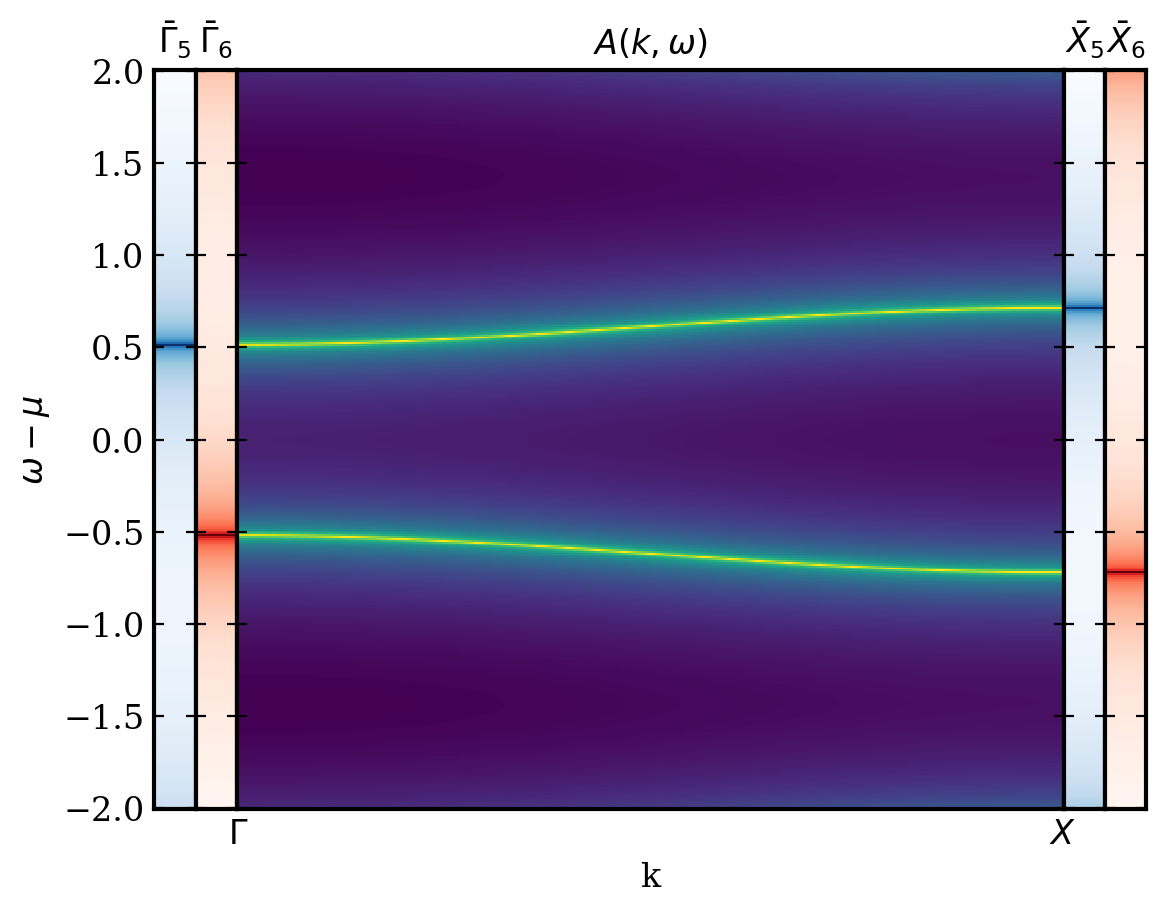}
        \caption{SAI $(\phi=\pi/4)$}
        \label{fig:SAI_spectral}%
    \end{subfigure}
    \par
    \begin{subfigure}[b]{0.45\linewidth}
        \centering
        \includegraphics[width=\linewidth]{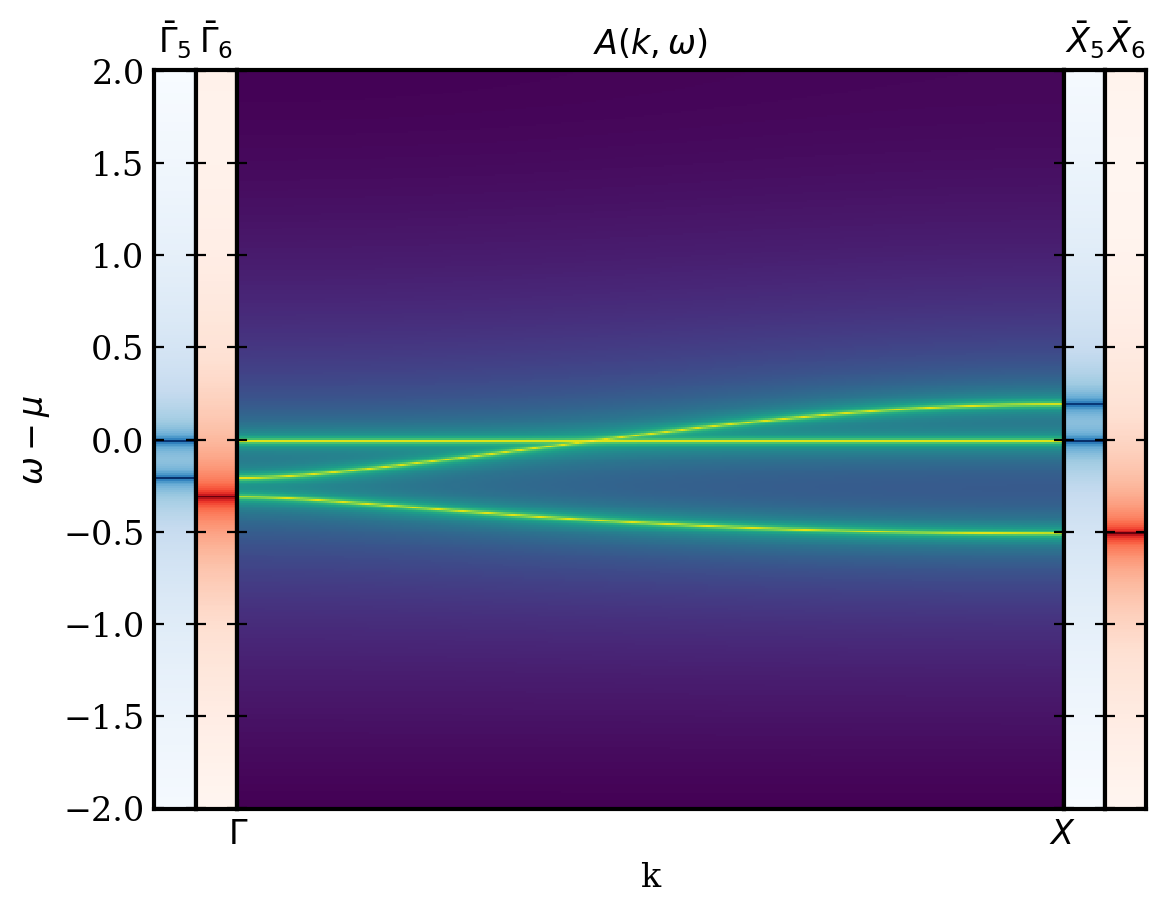}
        \caption{Metal-II $(\phi=(1/2 - 10^{-4})\pi)$}
        \label{fig:MetalII_spectral}
    \end{subfigure}
    \caption{Spectral functions of the three non-interacting phases directly associated with the interacting phases studied in subsection \ref{subsec:results_spectral}. Sidebars indicate the spectral function contributions by irrep.}
    \label{fig:non_interacting_spectral}
\end{figure}

\end{appendix}





\nocite{*}
\bibliography{ref_fixed_utf8_mended.bib}


\end{document}